\pdfoutput=1
\documentclass[11pt]{arxiv-article}
\bibliography{References}

\usepackage{Definitions}

\usepackage{hyperref}
\usepackage{mathrsfs}  
\usepackage{multirow}

\usepackage{tikz}
\newcommand{\OfficialTitle}{
  Large charge at large N
}
\title{\setstretch{1.4}
  {\color{Thoughtless}\Huge\textbf{\dosserif\OfficialTitle}}
}

\hypersetup{pdfauthor={Some people},pdftitle={\OfficialTitle}}

\author{%
  \begin{minipage}{.97\linewidth}
    \vspace{1cm}
    \begin{center} \dosserif%
      {\small
         \textbf{Luis~Alvarez-Gaume}\textsuperscript{\ding{71}\ding{95}},
         \textbf{Domenico Orlando}\textsuperscript{\ding{72}\ding{73}} and
         \textbf{Susanne Reffert}\textsuperscript{\ding{73}} 
         }
    \end{center}
    \vspace{1cm}
     \authorBlock{\ding{71}}{\dosserif{} Simons Center for Geometry and Physics,\\ State University of New York
       Stony Brook,\\
       NY--11794--3636, USA}
     \authorBlock{\ding{95}}{Theory Department -- CERN,\\ 
 \textsc{ch}-1211 Geneva 23, Switzerland}
     \authorBlock{\ding{72}}{\dosserif{} INFN sezione di Torino | Arnold--Regge Center\\
      via Pietro Giuria 1, 10125 Turin, Italy}
    \authorBlock{\ding{73}}{\dosserif{} Albert Einstein Center for Fundamental Physics\\
      Institute for Theoretical Physics, University of Bern,\\
      Sidlerstrasse 5, CH-3012 Bern, Switzerland}
  \end{minipage}
}

\date{}

\begin{document}

\setstretch{1.2}

\numberwithin{equation}{section}

\begin{titlepage}

  \newgeometry{top=23.1mm,bottom=46.1mm,left=34.6mm,right=34.6mm}

  \maketitle

  \thispagestyle{empty}

  \vfill\dosserif{}

  \abstract{\normalfont{}\noindent{}%
    We apply the large-charge expansion to \(O(N)\) vector models starting from first principles, focusing on the Wilson--Fisher point in three dimensions. We compute conformal dimensions at zero and finite temperature at fixed charge \(Q\), concentrating on the regime \(1 \ll N \ll Q\).
    Our approach places the earlier effective field theory treatment on firm ground and extends its predictions.
  }

\vfill

\end{titlepage}

\restoregeometry{}

\setstretch{1.2}

\tableofcontents

\section{Introduction}%
\label{sec:Introduction}

The study of quantum systems in the limit of large quantum numbers goes back
to the early years of quantum theory in terms of the \textsc{wkb} approximation.
In theories without an obvious expansion parameter, in many
\acp{cft} and in strongly coupled systems,  it is often useful to look
for special variables defining them, which under certain circumstances
allow a new approximation scheme for the physical problems of interest.
This is for example the case in the original semiclassical expansion in Quantum Mechanics, the
large-$N$ limit in \ac{qft}~\cite{Moshe:2003xn}\footnote{Yaffe~\cite{Yaffe:1981vf} discusses the sense in which large-$N$ limits
  of various quantum theories are equivalent  to classical limits.} and large-spin limits~\cite{Berenstein:2002jq,Komargodski:2012ek}.
Recently,
there has been a lot of interest in the study of \acp{qft} with
global symmetries in the limit of large charge.  For several \acp{cft}, the anomalous dimensions
of primary operators with large charge have been computed~\cite{Hellerman:2015nra,Alvarez-Gaume:2016vff,Monin:2016jmo,Loukas:2017lof,Orlando:2019hte}.

\bigskip

One of the aspects of the large-charge expansion that is particularly interesting
is its relation to the general study of symmetry breaking in non-relativistic
systems~\cite{guralnik1967broken,Nielsen:1975hm,Watanabe:2019xul,Brauner:2010wm,Watanabe:2013uya}.
This is a vast subject, which is probably far from being exhausted.  In the type
of theories we consider, the microscopic theory is a local relativistic theory,
and we are interested in the large-charge sector.  The type of non-relativistic
breaking of symmetry we find is closer to the spontaneous symmetry probing
explored in~\cite{Nicolis:2011pv,Nicolis:2013sga,Nicolis:2013lma}.  We will recover the same hybrid
pattern of symmetry breaking we found in~\cite{Alvarez-Gaume:2016vff} when studying vectors models,
but using large-$N$ methods as a way of dealing with the strong coupling
of the original theory.
\vskip.5cm
In~\cite{Hellerman:2015nra}, using an orthodox \ac{rg} approach to study the
$O(2)$-Wilson--Fisher fixed point theory in three-dimensions (among others), 
a number of universal features were uncovered at long distances in the large-charge
sector.  Among them is the existence of a type-I Nambu--Goldstone boson.  In the
nomenclature of~\cite{Nielsen:1975hm}, these are massless excitations whose low-energy
dispersion relations take the form, $\omega\sim p$.  In the case at hand
the precise relation is fixed by conformal invariance to be $\omega=p/\sqrt{d-1}=p/\sqrt{2}$ in three space-time
dimensions.  At long distances, the large-charge sector of the theory can 
be described in terms of a well-defined \ac{eft} for these
Goldstone bosons, where the interactions are suppressed by inverse powers
of the charge.  Furthermore, the anomalous dimension of a primary field of
charge $Q$ has the general form 
\begin{equation}
  \label{eq:DeltaQ}
  \Delta(Q) = c_{3/2} Q^{3/2} + c_{1/2} Q^{1/2}+
  c_0 + \order{Q^{-1/2}}.
\end{equation}
The coefficients $c_{3/2} $ and $c_{1/2}$ depend on the theory,
but the constant term  $c_0 \approx -0.0937254$ is universal and is a prediction of the
theory.\footnote{For the calculation via the zeta-function regularization, see~\cite{Monin:2016bwf}. This value was verified numerically in~\cite{delaFuente:2018qwv} in the context of the $CP^{N-1}$ model.}

\bigskip

In~\cite{Alvarez-Gaume:2016vff} the analysis in~\cite{Hellerman:2015nra} was extended to the $O(2N)$ vector theory.  We can
choose the maximal Abelian subgroup $O(2)^N$ as a set of independent 
rotations on the $N$ orthogonal $2$-planes inside $\setR^{2N}$, and put an
arbitrary charge $Q_i$ on each of them.  Let $\varphi_j, j=1,\ldots N$ be 
the complex variables describing the orthogonal two-planes where
each $O(2)$ acts as a phase rotation.  
The semiclassical analysis shows that in this case, the minimal energy state 
corresponds to uniform rotations in each of the two-planes, with the same
frequency $\varphi_j={ A_j\over \sqrt{2}} e^{i \mu t}$.  All two-planes share
the same clock (this is reminiscent of the spontaneous symmetry probing phenomenon in~\cite{Nicolis:2011pv}).  The
exponent $\mu$ depends on the details of the field theory, but for the
scale-invariant theory, 
$\mu\sim Q^{1/(d-1)}$, $Q= \sum_i \abs{Q_i}$.

\bigskip

By perturbing in the scaling theory around the minimal state with large
$Q_i$ charges in~\cite{Alvarez-Gaume:2016vff} we found that the low-energy theory contains
$(N-1)$ type-II Goldstone bosons. In the terminology of~\cite{Nielsen:1975hm}, they correspond
to excitations whose dispersion relation behaves like $\omega\sim p^2$ at
low energy, \emph{i.e.} they are Schrödinger particles. We find the same type-I
Goldstone boson as in~\cite{Hellerman:2015nra}, and of course we recover the same Casimir
energy contribution to the general form of the anomalous dimension
for the lowest-lying operator of charge $Q$.  It was also shown that the low-energy
limit of this $O(2N)$ model is described in terms of the Goldstone modes,
and the interaction terms in the Hamiltonian are suppressed by powers
of $Q$, in agreement with~\cite{Hellerman:2015nra}.

\bigskip

Using \ac{rg} arguments similar to those in~\cite{Hellerman:2015nra} we could have obtained the
scale-invariant theory in~\cite{Alvarez-Gaume:2016vff}, and then expanded around it.  In this paper, we take a different point of view, that will be less heuristic,
but will yield the same results and offer a simpler
procedure to compute corrections in the large-$Q$ expansion.

\bigskip

The idea is to start with a general-principle definition of the partition function
for a theory with well-defined global charge (see Sec.~\ref{sec:EFTlambda}).  We work on
$S^1_{\beta}\times S^2$,  as we want to make use of the state-operator correspondence of \ac{cft} (a similar approach to the study of the \(CP^{N-1}\) model at fixed monopole charge was used in~\cite{PhysRevB.78.214418,Pufu:2013eda,Dyer:2015zha}).
In this case, the low-temperature limit will be dominated
by the lowest-energy state carrying the charges imposed by the $\theta_i$
integrations.  For large charges  $Q_i$, the exponential 
prefactors oscillate very fast, and hence the leading
contribution to the partition function will be dominated by saddle points.
When analyzing the saddle point equations and their reliability, we first use 
zeta-function regularization to evaluate the path integral representing
the trace, and then we study the interplay between the large-$Q$ and
large-$N$ limits in different regimes.  We recover the expected results. The $1/Q$-expansion appears as a direct consequence of the heat-kernel expansion and is an asymptotic series in the technical sense.
The use of large $N$ allows us to compute the unknown coefficients in the anomalous dimension~\eqref{eq:DeltaQ}, finding encouraging agreement with the
lattice computations of $c_{3/2}$ and $c_{1/2}$~\cite{Banerjee:2017fcx,Banerjee:2019jpw}. The coefficient $c_0$ is of course
obtained with the same value.  

Using $S^1_{\beta}$ allows us to study the low-temperature regime, going beyond the earlier \ac{eft} results at $T=0$. It is interesting how
the saddle-point expansion together with the large-$N$ limit reorganize the
quantum expansion in such a way that we can easily recognize
how the interplay between large $Q$ and large $N$ gives rise to the various terms.  To discuss the properties of the
universal (conformal) Goldstone boson we need to study
the quantum structure at order $\order{N^0}$ in the auxiliary fields
used to represent the large-$N$ formulation of the theory
(Section~\ref{sec:goldstones}).  

A nice check of the self-consistency of our arguments is that by using these techniques, but at zero charges, we recover that the leading operator contributing is the identity operator at the conformal point.

\bigskip

The plan of this article is as follows. In Section~\ref{sec:action}, we study the action of the $O(2N)$ vector model in three dimensions at the infrared fixed point at fixed global charge.
We write down an effective action for the field $\hat\lambda$ which is promoted from Lagrange multiplier by integrating out the original fields $\varphi_i$.
$\hat\lambda$ describes the fluctuations around the saddle point of the path integral, which we study in Section~\ref{sec:saddle}, making use of the zeta-function regularization. We separately explore the regimes of zero charge and zero temperature (Section~\ref{sec:zero-Q-zero-T}), finite charge and zero temperature (Section~\ref{sec:large-Q-zero-T}) and finite charge and finite temperature (Section~\ref{sec:large-Q-finite-T}). While the results obtained this far can be interpreted as leading-order terms in $N$, we next identify the Goldstone spectrum by studying the relevant next-to-leading order contributions in Section~\ref{sec:goldstones}. We end with conclusions and outlook in Section~\ref{sec:conclusions} and collect the necessary technical details pertaining to the zeta-functions in Appendices~\ref{sec:S1-Sigma-zeta-function} and \ref{sec:sphere-zeta-function}. In Appendix~\ref{sec:one-loop-lambda-propagator}, the one-loop term in the propagator for \(\hat \lambda\) is calculated.

\section{The action}%
\label{sec:action}

\subsection{The infrared fixed point}%
\label{sec:infraredFP}

We start with the Landau--Ginzburg model for \(2N\) real scalar fields in the vector representation of \(O(2N)\) in \((1+2) \) dimensions with Euclidean signature on \(S^1_\beta \times \Sigma\), where $\Sigma$ is a Riemann surface.\footnote{Since we are ultimately interested in using the state-operator correspondence of conformal field theory, we will mostly work on $\Sigma=S^2$.}
Keeping all the terms up to mass dimension three, we have
\begin{equation}\label{eq:UV-Hamiltonian}
  S_\theta[\varphi_i] = \sum_{i=1}^N  \int \dd{t} \dd{\Sigma} \bqty{g^{\mu\nu} \pqty{\del_\mu \varphi_i}^* \pqty{\del_\nu \varphi_i} + r \varphi_i^* \varphi_i  + \frac{u}{2}  \pqty{\varphi_i^* \varphi_i}^2 + \frac{v}{4}  \pqty{\varphi_i^* \varphi_i}^3}
\end{equation}
where \(\varphi_i\) are complex fields.
  Note that the term \(r \varphi_i^* \varphi_i\) includes also the coupling to the scalar curvature of \(\Sigma\), which in our case is constant.
We will see that this model flows to the \ac{wf} in the \ac{ir} limit \(u \to \infty\) when \(r\) is fine-tuned to the value fixed by conformal coupling.
The \(\abs{\varphi}^6\) term is marginally irrelevant.

We are interested in the canonical partition function at fixed charge, where we fix the charges \(Q_i\) that act as rotations on  the complex fields \( \varphi_i \), \emph{i.e.} the Noether charges
\begin{equation}
  \hat Q_i = \int \dd{\Sigma} j_i^0 =  i \int \dd{\Sigma} \bqty{  \dot \varphi_i^* \varphi_i - \varphi_i^* \dot \varphi_i} .
\end{equation}
The partition function takes the form
\begin{equation}%
  \label{eq:canonicalPartitionFn}
  \begin{aligned}
  Z(Q_1, \dots, Q_N) &= \Tr[ e^{-\beta H} \prod_{i=1}^N \delta(\hat Q_i - Q_i)]\\
  & = \int_{-\pi}^{\pi} \prod_{i=1}^N \frac{\dd{\theta_i}}{2 \pi} \prod_{i=1}^N e^{i \theta_i Q_i} \Tr[ e^{-\beta H} \prod_{i=1}^N e^{- i \theta_i \hat Q_i}] .
\end{aligned}
\end{equation}
In this Hamiltonian representation, charge quantization implies that the eigenvalues of the $\hat Q_i$ are integers.
This in turn implies that the integrand is a $2\pi$-periodic function in each of the $\theta_i$, allowing us to use the integration bounds of $-\pi$ and $\pi$.

The integral over \(\theta_i\) can be solved in terms of an asymptotic expansion in \(1/Q_i\).
In general such an expansion will receive contributions from the end points of the integration \(\theta_i = \pm \pi\ \) and from the saddle points of the integrand.
In our case, however, the integrand is a \((2\pi)\)-periodic function of \(\theta_i\), so the contributions from the end points cancel each other and the leading contribution comes from the saddle point.

The trace describes the grand-canonical partition function for a theory with  imaginary chemical potentials \(\mu_i = i \theta_i/\beta \) associated to the currents \(j_i^0\).
The \(\theta\)-dependent terms break the original \(O(2N)\) symmetry to the \(U(N)\) that acts linearly on the complex fields \(\varphi_i\).
The saddle-point equation for \(\theta_i\) is $iQ_i + \del_{\theta_i} F_{gc}(i\theta/\beta) = 0$, where \(F_{\text{gc}}\) is the corresponding free energy.
  We will be discussing compact manifolds, so we expect \(F_{\text{gc}}\) to be smooth and the derivatives to be well-defined.  
If the theory is $CP$-invariant, \(F_{\text{gc}}\) is necessarily an even function of $\mu$.
The reason is that under $T$, the chemical potential transforms as $\mu \to - \mu$ and if the theory is $CP$ invariant, then $\mu \to  - \mu$ has a to be a symmetry. %
It follows that at the saddle point $\theta$ has to be necessarily imaginary, since the derivative of $F_{gc}(\mu)$ is an odd real function:
taking the complex conjugate of the equation one finds that at the minimum $\theta_i^* = - \theta_i$, and equivalently, the \(\mu_i\) are real.

\bigskip

Since the current \(j^0\) in the trace depends on the momenta, the sum over the momenta  is non-trivial.
The result can be understood in two equivalent ways:
\begin{itemize}
\item imposing a non-trivial \ac{bc} on the fields \(\varphi_i\) around the thermal circle:
  \begin{equation}
    \Tr[e^{-\beta H -i \theta \cdot \hat{Q}}] = \int\displaylimits_{\varphi_i( \beta, x) = e^{i \theta_i} \varphi_i(0,x)} \DD{\varphi_i} e^{-S[\varphi_i]}.
  \end{equation}
  Since the fields $\varphi_i$ are all scalars, shifting $\theta_i$ by $2\pi$ results in the same boundary condition.
  We see again that the trace is a $2\pi$-periodic function.
\item introducing a constant background field in the time direction for the gauged \(U(1)\) symmetry and keeping periodic \ac{bc}:
\begin{equation}
    \Tr[e^{-\beta H -i \theta \cdot \hat{Q}}] = \int\displaylimits_{\varphi_i(\beta, x) = \varphi_i(0,x)} \DD{\varphi_i}  e^{-S_\theta[\varphi_i]},
  \end{equation}
  where the gauged Euclidean action is
  \begin{equation}
    S_\theta[\varphi_i] = \sum_{i=1}^N  \int \dd{t} \dd{\Sigma} \bqty{g^{\mu\nu} \pqty{D_\mu^i \varphi_i}^* \pqty{D^i_\nu \varphi_i} + r \varphi_i^* \varphi_i  + \frac{u}{2}  \pqty{\varphi_i^* \varphi_i}^2 + \frac{v}{4}  \pqty{\varphi_i^* \varphi_i}^3}
  \end{equation}
  and the covariant derivative is
  \begin{equation}
    D^i_\mu \varphi_i = \begin{cases}
      \pqty{\del_0 + i \frac{\theta_i}{\beta} } \varphi_i & \text{if \(\mu = 0\)} \\
      \del_i \varphi & \text{otherwise.}
    \end{cases}
  \end{equation}
\end{itemize}

We are interested in the \ac{ir} behavior where the theory flows to a \ac{wf} point.
We will show that this is the case in the limit \(u \to \infty\) with \(r\) and \(v\) finite.
In the standard approach to the vector model~\cite{Moshe:2003xn,ZinnJustin:2007zz} we start with a Hubbard--Stratonovich transformation~\cite{stratonovich1957method,Hubbard:1959}.
Introduce a Lagrange multiplier \(\lambda\) and an auxiliary field \(\eta\):
\begin{equation}
  \int \DD{\varphi_i} e^{-S_\theta[\varphi_i]} =  \int \DD{\varphi_i} \DD{\lambda} \DD{\eta} e^{-S_\theta[\varphi_i, \lambda, \eta]},
\end{equation}
where
\begin{multline}
  S_\theta[\varphi_i, \lambda, \eta] = \sum_{i=1}^N  \int \dd{t} \dd{\Sigma} \Bigg[ \pqty{D_\mu^i \varphi_i}^* \pqty{D^i_\mu \varphi_i} + r \varphi_i^* \varphi_i \\-  \lambda \pqty{\eta - \pqty{ \varphi_i^* \varphi_i + \frac{v}{4 u} \pqty{\varphi_i^* \varphi_i}^2 }} 
  - \frac{u}{2} \eta^2\Bigg].
\end{multline}
Integrating out \(\lambda\) reproduces our action, up to an irrelevant \(\pqty{\varphi^* \varphi}^4\) term.
We can now trade \(\pqty{\varphi_i^* \varphi_i}\) and \(\pqty{\varphi_i^* \varphi_i}^2\) for \(\eta^2\) and integrate out the field \(\eta\) since it only appears quadratically.%
\footnote{%
    The contour of integration for the Lagrange multiplier is chosen so to reproduce the result of perturbation theory in terms of a simple semiclassical computation. In this sense the auxiliary fields are a form of parametrizing many possible potentials that are related to the same infrared fixed point. Here we follow standard practice~\cite{Zinn-Justin:572813}.  %
}
The result is a new action, without quartic or sextic interactions, that depends on \(\varphi_i \) and the (former) Lagrange multiplier \(\lambda\)
\begin{equation}
  \int \DD{\varphi_i} \DD{\lambda} \DD{\eta} e^{-S_\theta[\varphi_i, \lambda, \eta]} = \int \DD{\varphi_i} \DD{\lambda} e^{-S_\theta[\varphi_i, \lambda]}, 
\end{equation}
where now%
\begin{equation}
  S_\theta[\varphi_i, \lambda] = \sum_{i=1}^N \int \dd{t} \dd{\Sigma} \bqty{ \pqty{D^i_\mu \varphi_i}^* \pqty{D^i_\mu \varphi_i} + \pqty{r + \lambda } \varphi_i^* \varphi_i  - \frac{\lambda^2}{2 u} + \frac{v}{4u} \lambda \pqty{\varphi_i^* \varphi_i}^2 } .
\end{equation}
We are interested in the \ac{ir} behavior of the system at the conformal point where \(r\) is fine-tuned to remain of order \(\order{1}\), \(u\) diverges and \(v\) is generically of order \(\order{1}\) since it is dimensionless.
The last two terms are both irrelevant (they have mass dimension \(4\)) and can be dropped without changing the physics of the problem.\footnote{This is consistent with the observation in~\cite{Appelquist:1981sf,Appelquist:1982vd} that the \(\abs{\varphi}^6\) term is marginally irrelevant in presence of the \(\abs{\varphi}^4\) term.}
This is a clear example of universality.
The physics at the fixed point is precisely the same independently of the values of the parameters in the initial (\ac{uv}) action.

Since we are working at fixed charge, our problem has two intrinsic scales: the charge density \(\rho_i = Q_i/ V\) and the scale \(u\) that has mass dimension \(1\). We fix \(Q_i\) such that the hierarchy 
\begin{equation}
  \frac{1}{L} \ll \Lambda \ll \rho_i^{1/2} \ll u 
\end{equation}
is satisfied, where \(L\) is the typical scale of the surface \(\Sigma\), and \(\Lambda \) is the energy scale for the physics that we want to study.
In this regime we can safely approximate the action as
\begin{equation}
  S_Q = \sum_{i=1}^N \bqty{-i   \theta_i Q_i + \int \dd{t} \dd{\Sigma} \bqty{ \pqty{D^i_\mu \varphi_i}^* \pqty{D^i_\mu \varphi_i} + \pqty{r + \lambda } \varphi_i^* \varphi_i  }} 
\end{equation}
with $D_0^i= \del_0 +i \theta_i /\beta $.
The $N$ modes are not independent since they share the same coupling to \(\lambda\).
This is why the system has \(U(N)\) symmetry as opposed to \(U(1)^N\).

\subsection{The effective action for \(\lambda\)}%
\label{sec:EFTlambda}

The fields \(\varphi_i\) appear quadratically in the action, and we can --- at least formally --- integrate them out.
Their inverse propagator is
\begin{equation}
  \Delta^{-1} = \begin{pmatrix}
    0 & \pqty{\omega - \frac{\theta}{\beta} }^2 + p^2 + \pqty{r + \lambda}\\
    \pqty{\omega - \frac{\theta}{\beta} }^2 + p^2 + \pqty{r + \lambda} & 0
  \end{pmatrix},
\end{equation}
and it has zeros for
\begin{equation}
  \label{eq:non-goldstone-dispersions}
  \omega^2 + \pqty{ \sqrt{r + \lambda} \pm \mu}^2 + \pqty{ 1 \pm \frac{\mu}{\sqrt{r + \lambda}} } p^2 \mp \frac{\mu}{4 \pqty{r + \lambda}^{3/2}} p^4 + \dots = 0.
\end{equation}
Here we have introduced $\mu=i\theta/\beta$ since we know that on general grounds, $\theta$ at the saddle point is imaginary.
The inverse propagator is non-singular, unless \(\mu^2 = r + \lambda\).
In this case, the fields \(\varphi_i\) will in general have non-trivial zero modes that need to be treated non-perturbatively.
We will see that this situation can be interpreted in terms of Goldstone's theorem: a non-trivial \ac{vev} (zero mode) marks a spontaneous breaking of the global symmetry and the appearance  of massless \acp{dof} in the spectrum.
Knowing that there can be a zero-mode for the \(\varphi_i\), we decompose the fields into a constant part \(A_i\) plus orthogonal fluctuations \(u_i\):
\begin{align}
  \varphi_i &= \frac{1}{\sqrt{2}} A_i + u_i , & \ev{u_i} = 0.
\end{align}
The action becomes (up to a total derivative)
\begin{equation}
  S_Q = \sum_{i=1}^N \bqty{ -i   \theta_i Q_i + \int \dd{t} \dd{\Sigma} \bqty{\pqty{D^i_\mu u_i}^* \pqty{D^i_\mu u_i}+ \frac{\theta_i^2 A_i^2}{2\beta^2} 
  + \pqty{r + \lambda } \abs{\frac{A_i}{\sqrt{2}} + u_i}^2  }} .
\end{equation}
It is convenient to also separate the field \(\lambda\) into a constant part and orthogonal (spacetime-dependent) fluctuations:
\begin{align}
  \lambda &= \pqty{m^2 - r} + \hat \lambda , & \ev{\hat \lambda} &= 0.
\end{align}
Now we can perform the quadratic path integral over the \(u_i\):
\begin{equation}
  Z(Q_1, \dots, Q_N) = \int \prod_{i=1}^N \frac{\dd{\theta_i}}{2\pi} \DD{u^i} \DD{\lambda} e^{-S_Q[u^i, \lambda]} = \int \prod_{i=1}^N \frac{\dd{\theta_i}}{2\pi} \DD{\hat \lambda} \prod_{i=1}^N e^{i \theta_i Q_i} e^{-S_{\theta}[\hat \lambda]},
\end{equation}
where
\begin{equation}
  S_{\theta}[\hat \lambda] =  \sum_{i=1}^N \bqty{  V \beta  \pqty{\frac{\theta_i^2}{\beta^2} + m^2 } \frac{A_i^2}{2} +  \Tr[\log(-D_\mu^i D_\mu^i + m^2 
    + \hat \lambda)] 
    - \frac{A_i^2 }{2} \Tr( \hat \lambda \Delta \hat \lambda) }.
\end{equation}
\(\Tr(\hat \lambda \Delta^i \hat \lambda )\) is a non-local term,
\begin{equation}
  \Tr(\hat \lambda \Delta^i \hat \lambda) = \int \dd{t} \dd{\Sigma} \dd{t'} \dd{\Sigma'} \hat \lambda(t,x) \Delta^i(t-t', x-x') \hat \lambda(t', x') ,
\end{equation}
written in terms of the propagator \(\Delta^i(t,x)\) for the field \(u^i\):
\begin{equation}
  \pqty{-D^i_\mu D^i_\mu + m^2} \Delta^i(r) = \frac{1}{\sqrt{g}} \delta(r).
\end{equation}
We can now separate the zero-modes in the action, expanding the functional determinant as a formal series:
\begin{multline}
  \label{eq:expanded-tr-log-original}
  S_{\theta}[\hat \lambda] =  \sum_{i=1}^N \Bigg[ V \beta \pqty{\frac{\theta_i^2}{\beta^2} + m^2 } \frac{A_i^2}{2} +  \Tr[\log(-D_\mu^i D_\mu^i + m^2)] - \frac{A_i^2 }{2} \Tr(\hat \lambda \Delta^i \hat \lambda) \\
  - \sum_{n=2}^\infty \frac{(-1)^n}{n } \Tr(\Delta^i \hat \lambda)^n \Bigg] ,
\end{multline}
where \(\Delta\hat \lambda\) are non-local terms written in terms of the propagator as
\begin{align}
  \Tr(\Delta^i \hat \lambda)^n &= \int \dd{r_1} \dots \dd{r_n}  \hat \lambda(r_1) \dots  \hat \lambda(r_n) \prod_{i } \Delta^i(r_i - r_{i-1}), & r_0 &= r_n .
\end{align}
Our final result is an effective action for the fluctuations \(\hat \lambda\) which depends on the parameters \(m^2\), \(A_i\) and \(\theta_i\).
We have redefined the quantum structure of the problem: instead of our original fields $\varphi_i$, we now have an action in terms of $\hat\lambda$, which has been promoted to a collective field from a Lagrange multiplier. The information about the fixed point and the symmetry-breaking pattern is contained in the zero-mode $m^2$, which acts as a \ac{rg} flow parameter.

\section{The saddle point}%
\label{sec:saddle}

In the previous section we have found an effective action for the field \(\hat \lambda\).
It describes the fluctuations around the saddle point of the path integral.\footnote{%
The integration over the $\theta_i$ angles are treated in terms of asymptotic expansions. The leading contributions come from saddle points in the complex plane since the integrand is \(2\pi\)-periodic and the contributions at \(\theta_i = \pm \pi\) cancel each other~\cite{Erdelyi:1956asymptotic}.%
}
In this section we will find the saddle point as a function of the control parameters \(Q\) and \(\beta\) and discuss the behavior of the system in the limit where the fluctuations can be neglected.
We will see in the following that this assumption is consistent in the large \(N\) limit.

\subsection{Saddle point equations}%
\label{sec:saddle-equations}

The saddle point equations are obtained deriving the effective action with respect to  \(m^2\), \(A_i\) and \(\theta_i\) at \(\hat \lambda = 0\):
\begin{equation}
  \label{eq:saddle-equations-Tr-log}
  \begin{dcases}
    \pdv{S_Q}{m^2} =  \sum_{i=1}^N \bqty{V \beta \frac{A_i^2}{2} + \pdv{m^2} \Tr[ \log( - D^i_\mu D^i_\mu + m^2)]} = 0 ,\\
    \pdv{S_Q}{\theta_i} = \frac{\theta_i}{\beta} V A_i^2 +  \pdv{\theta_i} \Tr[ \log( - D^i_\mu D^i_\mu + m^2)] - i Q_i = 0 , & i = 1,\dots, N\\
    \pdv{S_Q}{A_i} = V \beta \pqty{\frac{\theta_i^2}{\beta^2} + m^2 } A_i = 0 , & i = 1,\dots, N.
  \end{dcases}
\end{equation}

The first equation relates \(m^2\) to the quantum effects summed up in the functional determinant.
The value taken at the saddle point contains information about the dimensionality and the symmetries of the problem.
We will use this value at zero charge to verify that we sit indeed at the conformal point, while at large charge it will be the controlling parameter of the asymptotic expansion.
In this sense, \(m^2\) plays the same role as the (leading term of the) effective potential in the \ac{eft}.

The second equation shows how the charges \(Q_i\) are distributed between the \ac{vev} \(A_i\) and the fluctuations \(u^i\) that are summed in the functional determinant.
We will see explicitly that for \(\beta \to \infty\) the fluctuation part vanishes and all the charge  comes from  the \ac{vev}.

The third equation shows that a non-vanishing \ac{vev} is only possible if \(\theta_i^2 = - m^2 \beta^2\).
This is precisely the same condition \(\mu^2 = r + \lambda = m^2\) that we had found above for the \(\varphi_i\) to contain massless modes.

\bigskip

If we neglect the fluctuations  \(\hat \lambda \), we can identify the functional determinant with the grand-canonical (fixed chemical potential) free energy:
\begin{equation}
  F^{\saddle}_{gc}(i \theta) = \sum_{i=1}^N \bqty{ V \pqty{\frac{\theta_i^2}{\beta^2} + m^2 } \frac{A_i^2}{2} +  \frac{1}{\beta}\Tr[\log(-D_\mu^i D_\mu^i + m^2)]}.
\end{equation}
The canonical (fixed charge) free energy is then
\begin{equation}
  F^{\saddle}_{c} (Q) = \eval{-i \sum_{i=1}^N \frac{\theta_i}{\beta} Q_i + F^{\saddle}_{gc}(i \theta)}_{Q_i = -i\beta \pdv{\theta_i} F_{gc}(\theta)} = \eval{- \sum_{i=1}^N \mu_i Q_i + F^{\saddle}_{gc}(\mu)}_{Q_i = \pdv{ F_{gc}}{\mu_i}} ,
\end{equation}
where the value of \(\theta\) is fixed by the saddle point condition.
This is a non-trivial consistency check of our construction: at the saddle point we reproduce the usual Legendre transform that relates the two thermodynamic potentials.

\subsection{Zeta-function regularization}%
\label{sec:zeta-function-saddles}

The functional determinant that appears in Eq.~\eqref{eq:saddle-equations-Tr-log} needs to be regularized.
In view of using the state/operator correspondence, we are interested in the theory living on a two-sphere. It is therefore convenient to use the zeta function regularization as it does not affect the compactification manifold (see~\cite{Elizalde:2012zza,kirsten2001spectral} for an introduction).

In the zeta function scheme we introduce the sum over the eigenvalues of the operator \(-D_\mu D^\mu + m^2 \),
\begin{equation}
  \zeta(s | \theta, \Sigma, m) = \sum_{n \in \setZ} \sum_{p}  \pqty{ \pqty{\frac{2 \pi n}{\beta} + \frac{\theta}{\beta} }^2 + E(p)^2 + m^2}^{-s} .
\end{equation}
The \(E^2(p)\) are the eigenvalues of the Laplacian on \(\Sigma\),
\begin{equation}
  \Laplacian f_p(x) + E(p)^2 f_p (x) = 0 ,
\end{equation}
and we have used the fact that the \(\theta\)-terms shift the Matsubara frequencies on the thermal circle from \(2 \pi n/\beta\) to  \(2\pi n /\beta + \theta/ \beta\).

Then we write the functional determinant as
\begin{equation}
  \Tr[\log(- D_\mu D^\mu + m^2)] =  \sum_{n \in \setZ} \sum_{p} \log( \pqty{\frac{2 \pi n}{\beta} + \frac{\theta}{\beta} }^2 + E(p)^2 + m^2)
\end{equation}
and its derivatives that appear in the saddle point equations are:
\begin{equation}
  \begin{aligned}
    \pdv{m^2} \Tr[\log(- D_\mu D^\mu + m^2)] &= \sum_{n \in \setZ} \sum_{p}  \pqty{ \pqty{\frac{2 \pi n}{\beta} + \frac{\theta}{\beta} }^2 + E(p)^2 + m^2}^{-1} \\
    & = \eval{\zeta(s | \theta, \Sigma, m)}_{s = 1},
  \end{aligned}
\end{equation}
\begin{equation}                          
  \begin{aligned}
    \pdv{\theta} \Tr[\log(- D_\mu D^\mu + m^2)] &= \frac{2}{\beta} \sum_{n \in \setZ} \sum_{p} \frac{\frac{2 \pi n}{\beta} + \frac{\theta}{\beta}}{\pqty{\frac{2 \pi n}{\beta} + \frac{\theta}{\beta} }^2 + {E(p)}^2 + m^2} \\
    &= - \eval{ \frac{1}{s} \pdv{\theta} \zeta(s | \theta, \Sigma, m)}_{s = 0} .
\end{aligned}
\end{equation}
Collecting the above results, we can write the zeta-function-regulated saddle point equations:
\begin{equation}%
  \label{eq:saddle}
  \begin{dcases}
    \pdv{S_Q}{m^2} =  \sum_{i=1}^N \bqty{\frac{V \beta}{2} A_i^2 +  \zeta(1 | \theta_i, \Sigma, m)} = 0, \\
    \pdv{S_Q}{\theta_i} = -i Q + \frac{\theta_i}{\beta} V A_i^2 +  \eval{ \frac{1}{s} \pdv{\theta_i} \zeta(s | \theta_i, \Sigma, m)}_{s = 0} = 0  , & i=1,\dots, N\\
     \pdv{S_Q}{A_i} = V \beta \pqty{\frac{\theta_i^2}{\beta^2} + m^2 } A_i =0 , & i=1,\dots, N.
   \end{dcases}
\end{equation}
Finally, it is also convenient to write the \(\Tr \log(\cdot)\) term in this regularization, using the identity \(\log(x) = - \eval{\dv{x^{-s}}{s}}_{s= 0}\):
\begin{multline}
  \Tr[\log(- D_\mu D^\mu + m^2)] = - \eval{ \dv{s} \sum_{n \in \setZ} \sum_{p} \pqty{\pqty{\frac{2 \pi n}{\beta} + \frac{\theta}{\beta} }^2 + E(p)^2 + m^2 }^{-s} }_{s=0} \\ = - \eval{\dv{s} \zeta(s| \theta, \Sigma, m)}_{s=0}.
\end{multline}

\paragraph{A technical remark.}

In the following, we will use Mellin's representation of the zeta function:
\begin{equation}
  \zeta(s | \theta,  \Sigma, m) = \frac{1}{\Gamma(s)} \int_0^\infty \frac{\dd{t}}{t} t^s e^{-m^2 t } \sum_{n \in \setZ} e^{-\pqty{\frac{2 \pi n}{\beta} + \frac{\theta}{\beta}  }^2 t } \Tr[e^{\Laplacian_\Sigma{} t}].
\end{equation}
This has a number of advantages from our point of view.
In the large-charge limit, we expect \(m\) to be large for dimensional reasons.
In this case it will act as a regulator for the integral which, for \(m \gg 1\), localizes around \(t = 0\).
The integrand separates the contribution of the modes on the thermal circle clearly from the contribution of the internal manifold \(\Sigma\).
The former is a theta function which has a simple behavior for \(t \to 0\) and is manifestly \(2\pi\) invariant in \(\theta\) (see the Poisson-resummed form in Eq.~\eqref{eq:Poisson-resummation}); the latter can be expanded for small \(t\) using Weyl's asymptotic formula in Eq.~\eqref{eq:Weyl-asymptotic-formula}~\cite{weyl1911asymptotische}.
We use these properties extensively in the appendix to compute special values of the zeta functions in different limits.

\subsection{Zero charge, zero temperature: the conformal coupling}%
\label{sec:zero-Q-zero-T}

Let us stop a moment to examine the case without fixing the charge. Here, the main contribution comes from $Q=0$.
We can use this point as a litmus test to see whether the limit \(u \to \infty \) that we have taken indeed leads to a \ac{cft}.
We study the special case of zero temperature on the sphere, \(\beta \to \infty\) and  \(\Sigma = S^2\). At the fixed point, \(m^2\) must reproduce the coupling of a conformal massless scalar: 
\begin{equation}\label{eq:conf_coup}
	m^2 = \xi R = \frac{1}{8} \times \frac{2}{r_0^2}, 
\end{equation}
where $R$ is the Ricci scalar of $\Sigma$ and $r_0$ the radius of the two-sphere.
Moreover, the free energy must vanish, 
\begin{equation}
	F(Q=0) = 0,
\end{equation}
as by the state/operator correspondence it corresponds to the conformal dimension of the lowest operator of charge zero, which is the identity:
\begin{equation}\label{eq:conf_dim_Id}
	F(0) = r \Delta(\Id) = 0.
      \end{equation}
      
For \(\beta \to \infty\), the zeta function \(\zeta(s| \theta, \Sigma, m)\) does not depend on \(\theta\), and is proportional to the zeta function on the compactification manifold \(\Sigma\).
To see this, one can either start from the expression for finite \(\beta\) and take the limit  (see Eq.~(\ref{eq:zeta-for-beta-infinity})), or compute directly the zeta function summing over the eigenvalues on the (infinite) thermal circle,
\begin{multline}
  \label{eq:zeta-beta-infinity}
  \lim_{\beta \to \infty } \frac{1}{\beta}  \zeta(s | \theta, \Sigma, m) =  \frac{1}{2\pi} \int_{-\infty}^{\infty} \dd{\omega } \sum_p \pqty{ \pqty{\omega + i \mu  }^2 + E(p)^2 + m^2}^{-s} \\
  = \frac{\Gamma(s-\frac{1}{2})}{2 \sqrt{\pi} \Gamma(s)} \sum_p \pqty{E(p)^2 + m^2}^{1/2 - s} =\frac{\Gamma(s-\frac{1}{2})}{2 \sqrt{\pi} \Gamma(s)}  \zeta(s-\tfrac{1}{2}| \Sigma, m),
\end{multline}
where the sum runs over the eigenvalues \(E(p)^2\) of the Laplacian on \(\Sigma\).
For the particular values that we need in the saddle point equation and for the free energy, we get
\begin{align}
  \zeta(1 | \theta, \Sigma, m) &\xrightarrow[\beta\to \infty]{} \frac{\beta}{2} \zeta(\tfrac{1}{2}| \Sigma, m), \\
  - \eval{\dv{s} \zeta(1 | \theta, \Sigma, m)}_{s\to 0} &\xrightarrow[\beta\to \infty]{} \beta \zeta(-\tfrac{1}{2}| \Sigma, m).
\end{align}

If we do not fix the charge, there is no integration over the \(\theta_i\) and the saddle point equations are
\begin{equation}
  \begin{cases}
    \displaystyle \frac{\beta}{2} \sum_{i=1}^N \bqty{V A_i^2 +  \zeta(\tfrac{1}{2}| S^2, m)} = 0 ,\\
    V \beta m^2 A_i = 0, & i = 1, \dots, N.
  \end{cases}
\end{equation}
Since there are no external scales, based on naturalness we expect $m^2$ to be of order \(\order{1}\) and the appropriate expansion of the zeta function to be the one  given in Eq.~\eqref{eq:zeta-S2-small-m} in terms of Hurwitz zeta functions,
\begin{equation}
  \zeta(s |S^2, m) = 2 \sum_{k=0}^{\infty} \binom{-s}{k} \zeta(2s + 2k - 1, \tfrac{1}{2} )  \pqty{m^2 - \frac{1}{4} }^{k} .
\end{equation}
The important observation is that this is a Taylor series with constant term \(2 \zeta(2s-1, \tfrac{1}{2}) \) which vanishes both for \(s = 1/2\) and \(s=-1/2\).

For \(s = 1/2\), we have
\begin{equation}
    \zeta(\tfrac{1}{2} | S^2, m ) = - \frac{\pi^2}{2} \pqty{m^2 - \frac{1}{4} } + \frac{\pi^4}{8} \pqty{m^2 - \frac{1}{4} }^{2} + \dots 
\end{equation}
so the solution to the saddle point equation is
\begin{equation}
  \begin{cases}
    m^2 = \frac{1}{4}, \\
    A_i = 0,
  \end{cases}
\end{equation}
reproducing as hoped the conformal coupling~\eqref{eq:conf_coup}.

As expected, there are no zero modes for the fields \(\varphi_i\) and the information about the conformal point is entirely contained in the value of \(m^2\).

If we neglect the fluctuations \(\hat \lambda\), the free energy of the system is given by the value of the action at the saddle, which in our scheme depends on the zeta function in \(s=-1/2\):
\begin{equation}
  \zeta(-\tfrac{1}{2} | S^2, m ) = - \frac{\pi^2}{8} \pqty{m^2 - \frac{1}{4} }^{2} + \frac{\pi^4}{96} \pqty{m^2 - \frac{1}{4} }^{3} + \dots  
\end{equation}
and vanishes:
\begin{equation}
  F^{\saddle}(0) = - \frac{1}{\beta} \eval{\log(Z(Q))}_{Q=0} \approx  \eval{V \frac{A_i^2}{2} +  \zeta(-\tfrac{1}{2}|S^2, m)}_{m^2 = \frac{1}{4},  A_i = 0 } = 0 ,
\end{equation}
as predicted by the identity in Eq.~\eqref{eq:conf_dim_Id}.

\subsection{Finite charge, zero temperature: the broken phase}%
\label{sec:large-Q-zero-T}

Now that we have verified that we are in fact at the conformal point, we can study the case of \(Q \neq 0\), \(T = 0\) which should reproduce the \ac{eft} predictions of~\cite{Hellerman:2015nra,Alvarez-Gaume:2016vff}.

Using the low-temperature limit of the zeta function in Eq.~\eqref{eq:zeta-beta-infinity}, the saddle point equations reduce to
\begin{equation}
  \label{eq:saddle-finite-Q-no-T}
  \begin{cases}
    \frac{1}{2} \beta V v^2 + \frac{N \beta}{2}  \zeta(\tfrac{1}{2}| \Sigma, m) = 0, \\
    i Q_i - \frac{\theta_i}{\beta} V A_i^2 = 0, & i = 1, \dots, N,\\
    2 V \beta \pqty{m^2 + \frac{\theta_i^2}{\beta^2}} A_i = 0, & i = 1, \dots, N,
  \end{cases}
\end{equation}
where we have used the fact that the zeta function does not depend on \(\theta\) and we have introduced
\begin{equation}
  v^2 = \sum_{i=1}^N A_i^2. 
\end{equation}
From the second equation we see that if \(Q_i > 0\), \(A_i\) cannot vanish: at zero temperature, the charge is completely contained in the zero modes of the fields \(\varphi_i\).
From the third equation, we see that all the \(\theta_i\) are equal and take precisely the value \(\theta_i = \theta = i m \beta\) where the \(\varphi_i\) contain massless modes. This is consistent with our observations in Section~\ref{sec:EFTlambda} about $\theta_i$ being imaginary at the saddle point and the existence of zero modes.
We are in a broken phase, and we will see in the next section that the massless modes can be interpreted as Goldstone fields arranged in representations of the unbroken \(U(N-1)\) group.

The zero modes take the values
\begin{equation}
  A_i^2 = \frac{Q_i}{m V} 
\end{equation}
and the remaining equations are
\begin{equation}
  \begin{cases}
    m \zeta(\tfrac{1}{2} | \Sigma, m) = - \frac{Q}{N}, \\
    v^2 = \frac{Q}{m V},
  \end{cases}
\end{equation}
where
\begin{equation}
  Q = \sum_{i = 1}^N Q_i .
\end{equation}
As predicted in~\cite{Alvarez-Gaume:2016vff}, the saddle point depends only on the sum of the charges \(Q_i\). This was to be expected, as the problem has a $U(N)$ symmetry which can be used to rotate all the charges and the only invariant under this rotation is their sum.
In the \ac{eft}, this appears as a property of the homogeneous ground state at fixed charge.

\bigskip
If we neglect the fluctuations \(\hat \lambda\), the corresponding free energy is then given by\footnote{The contribution from the zero modes to the energy vanishes at the saddle point.}
\begin{equation}
  F^{\saddle}(Q) = - \frac{1}{\beta}  \sum_{i =1}^N \bqty{ i \theta_i Q_i - \beta \zeta(-\tfrac{1}{2}| \Sigma, m)} = m Q + N \zeta(-\tfrac{1}{2}| \Sigma, m),
\end{equation}
where  \(m\) is the saddle point value.

Our final result for the free energy at fixed charge \(Q\) at leading order in the fluctuations of \(\lambda\) on \(\setR \times \Sigma\) is
\begin{equation}
  \label{eq:result-finite-Q-zero-T-zeta}
  \begin{cases}
    F^{\saddle}_\Sigma(Q) = m Q + N \zeta(-\tfrac{1}{2}| \Sigma, m) , \\
    m \zeta(\tfrac{1}{2}| \Sigma, m) = - \frac{Q}{N} .
  \end{cases}
\end{equation}
The natural parameter that appears is \(Q/N\) which we hold fix.
In the following we will study the systems in the limit of \(Q/N \gg 1\) in order to express the zeta functions as asymptotic expansions.

\bigskip

We use Mellin's representation of the zeta function,
\begin{equation}
  \zeta(s |  \Sigma, m) = \frac{1}{\Gamma(s)} \int_0^\infty \frac{\dd{t}}{t} t^s e^{-m^2 t } \Tr[e^{\Laplacian_\Sigma{} t}].
\end{equation}
In the limit of large \(Q/N\), we expect \(m\) to be parametrically large, so we can use the corresponding asymptotic expansion where the integral localizes around \(t \to 0\).
The trace over the eigenvalues of the Laplacian is expressed using Weyl's asymptotic formula in terms of heat kernel coefficients:
\begin{equation}
  \Tr(e^{\Laplacian_\Sigma{} t}) = \sum_{n=0}^{\infty} K_n t^{n/2 - 1} .
\end{equation}
The heat kernel coefficients \(K_n\) can be computed using geometric invariants of the surface \(\Sigma\) (see~\cite{rosenberg_1997,Vassilevich:2003xt} for a detailed introduction) and one finds that if \(\Sigma\) has no boundary, all the odd coefficients vanish, \(K_{2n+1} = 0\).
The leading coefficients are given in terms of the volume \(V\) and the scalar curvature \(R\) of \(\Sigma\),
\begin{align}
  K_0 &= \frac{V}{4 \pi}, & K_2 &= \frac{V R}{24 \pi} . 
\end{align}
Each consecutive order in the large-charge expansion is obtained by taking the next term in the heat kernel expansion.

Let us explicitly consider some special cases for the manifold $\Sigma$.
\paragraph{The torus \(\Sigma = T^2\).}
The only non-vanishing heat kernel coefficient is \(K_0\). So, up to exponential corrections of order \(\order{e^{-m}}\),  we can use Eq.~(\ref{eq:zeta-T2-zero-T}):
\begin{equation}
  \zeta(s | T^2, m) = \frac{V}{4 \pi \pqty{s - 1}} m^{2-2s} .
\end{equation}
At the saddle point,
\begin{align}
  \label{eq:torus-energy-zero-T}
  m_{T^2}(Q) &= \sqrt{\frac{4 \pi}{V} } \pqty{\frac{Q}{2N} }^{1/2} , \\
  \frac{F^{\saddle}_{T^2}(Q)}{2N} &= \frac{2}{3} \sqrt{\frac{4\pi}{V} } \pqty{\frac{Q}{2N} }^{3/2} .
\end{align}
As expected, in the limit \(Q/N \gg 1\), at the saddle point  \(m \gg 1 \).

\bigskip

There is an independent verification of this result.
In~\cite{Appelquist:1982vd}, the authors compute a one-loop effective action for the \(O(2N)\) model, and they find that at leading order in \(N\) there is an effective potential,%
\begin{equation}
  V_{\text{1-loop}}(\varphi ) = \frac{16 \pi^2}{3N^2} \pqty{\varphi^i \varphi^i}^3.
\end{equation}
This is  the same type of effective potential considered in~\cite{Alvarez-Gaume:2016vff}.
Matching the conventions we find that the leading contribution to the energy of the homogeneous ground state of the effective action agrees precisely with the free energy that we have computed,
\begin{equation}
  E_{\text{1-loop}} = \frac{2 \sqrt{2 \pi}}{3 \sqrt{N V}}  Q^{3/2} .
\end{equation}
\paragraph{The unit sphere \(\Sigma = S^2\).}

To study the \(Q/(2N) \gg 1\) limit, we can use the explicit integral representation of the heat kernel~\cite{McKean:1967xf}:
\begin{equation}
  \Tr[e^{\Laplacian_{S^2}{} t}]  = 2 \frac{e^{t/4}}{\sqrt{\pi} t^{3/2}} \int_0^1 \dd{y} y \frac{e^{-y^2/t}}{\sin(y)}.
\end{equation}
Using it we can find an integral representation of the zeta function (see Eq.~\eqref{eq:zeta-S2-integral-representation}). 
The expression is quite involved but one can show that there is a natural  asymptotic expansion in terms of the mass of a conformally coupled scalar \((m^2 - 1/4)\). In Appendix~\ref{sec:sphere-zeta-function} we show that there is an optimal truncation at the fourth term,
\begin{multline}
  \zeta(s | S^2, m) = \frac{1}{s - 1} \pqty{m^2 - \frac{1}{4}}^{1-s} + \frac{1}{12} \pqty{m^2 - \frac{1}{4}}^{-s} + \frac{7 s}{480} \pqty{m^2 - \frac{1}{4}}^{-1-s} \\
  + \frac{31 s (s+1)}{8064} \pqty{m^2 - \frac{1}{4}}^{-2-s} \dots
\end{multline}
We can then express \(m\) and \(F\) as asymptotic expansions in \(1/Q\):
\begin{align}
  m_{S^2}(Q) &= \pqty{\frac{Q}{2N} }^{1/2} + \frac{1}{12} \pqty{\frac{Q}{2N} }^{-1/2} + \frac{7}{1440} \pqty{\frac{Q}{2N} }^{-3/2} + \frac{71}{120960} \pqty{\frac{Q}{2N} }^{-5/2} + \dots \\
  \label{eq:conformal-dimensions-large-Q}
  \frac{F^{\saddle}_{S^2}(Q)}{2N} &= \frac{2}{3} \pqty{\frac{Q}{2N} }^{3/2} + \frac{1}{6} \pqty{\frac{Q}{2N} }^{1/2} - \frac{7}{720} \pqty{\frac{Q}{2N} }^{-1/2} - \frac{71}{181440} \pqty{\frac{Q}{2N} }^{-3/2} +\dots
\end{align}
By the state-operator correspondence, this last expression is also the conformal dimension of the lowest operator of charge \(Q\) for the theory on \(\setR^3\),
\begin{equation}
  \Delta(Q) = r F_{S^2}(Q).
\end{equation}
These conformal dimensions were also computed on the lattice for \(N=1\) and \(N = 2\) in~\cite{Banerjee:2017fcx,Banerjee:2019jpw}.
Although this is far from the regime of validity of our approximation, the values are definitely in the same ballpark and for the \(O(4)\) model the discrepancy is of the order of \(10\%\), see table~\ref{tab:lattice-comparison} for details.

\begin{table}
  \centering
      \begin{tabular}{ccrrr}
      \toprule
      && lattice & leading \(N\) & error\\
      \midrule
      \multirow{2}{*}{\(O(2)\)} & \(c_{3/2}\) & \(0.337\) & \(0.471\) & \(40\%\) \\
      & \(c_{1/2}\) & \(0.266\) & \(0.236\) & \(10\%\) \\
      \multirow{2}{*}{\(O(4)\)} & \(c_{3/2}\) & \(0.301\) & \(0.333\) & \(10\%\) \\
      & \(c_{1/2}\) & \(0.294\) & \(0.333\) & \(13\%\)\\
      \bottomrule
    \end{tabular}

  \caption{Lattice results for the coefficients  \(c_{3/2}\) and \(c_{1/2}\) in the large-charge expansion of the conformal dimension \(\Delta(Q) = c_{3/2}Q^{3/2} + c_{1/2} Q^{1/2} + \dots \) in the \(O(2)\) and \(O(4)\) models~\cite{Banerjee:2017fcx,Banerjee:2019jpw}, compared with the large-\(N\) result in Eq.~\eqref{eq:conformal-dimensions-large-Q}.}%
  \label{tab:lattice-comparison}
\end{table}

\bigskip

The saddle equations~\eqref{eq:result-finite-Q-zero-T-zeta} are valid for any value of the charge.
So we can also use them in the opposite limit of \(Q/N \ll 1\), where the correct asymptotic expansion of the zeta function on the sphere is the one in Eq.~\eqref{eq:zeta-S2-small-m}.
In this approximation we are close to the zero-charge case discussed in Section~\ref{sec:zero-Q-zero-T}.
Expanding at \ac{nlo} we find that \(m\) receives a \(Q/N\) correction to the conformal coupling value
\begin{equation}
  m = \frac{1}{2} + \frac{8}{\pi^2} \frac{Q}{2N} + \order{\pqty{\frac{Q}{2N} }^2},
\end{equation}
and the corresponding conformal dimension is
\begin{equation}
  \label{eq:conformal-dimension-small-charge}
  \Delta(Q) = N \pqty{\frac{Q}{2N}  + \frac{8}{\pi^2} \pqty{\frac{Q}{2N} }^2 + \order{\pqty{\frac{Q}{2N} }^3} }.
\end{equation}
The leading term was to be expected because in this limit we can identify the lowest operator of charge \(Q\) with \(\varphi^Q\), which has engineering dimension \(Q/2\).

\paragraph{Generic Riemann surface.}
Each term in the heat kernel expansion corresponds to a term in the large-charge expansion of the free energy.
For a generic Riemann surface \(\Sigma\) without boundaries, the odd heat kernel coefficients vanish and the expansion is in integer powers of \(1/Q\) (as opposed to \(1/Q^{1/2}\)).
Since the expansion starts at \(Q^{3/2}\), there are only two terms that are not suppressed at large charge in the expansion of \(F(Q)\).
To compute them, we only need to keep the first two terms in Weyl's asymptotic expansion.
For a surface of volume \(V\) and scalar curvature \(R\) we find
\begin{align}
  m_\Sigma &= \sqrt{\frac{4\pi}{V} } \pqty{\frac{Q}{2N} }^{1/2} + \frac{R}{24} \sqrt{\frac{V}{4\pi} } \pqty{\frac{Q}{2N} }^{-1/2} + \dots \\
  \label{eq:free-energy-zero-T-any-sigma}
  \frac{F^{\saddle}_\Sigma}{2N} &= \frac{2}{3} \sqrt{\frac{4\pi}{V} } \pqty{\frac{Q}{2N} }^{3/2} + \frac{R}{12} \sqrt{\frac{V}{4\pi}} \pqty{\frac{Q}{2N} }^{1/2} + \dots
\end{align}
This formula reveals the existence of a $Q$-expansion, whose coefficients contain a manifold-dependent part and a model-dependent part.
The result is in complete agreement with the predictions of the \ac{eft} of~\cite{Hellerman:2015nra,Alvarez-Gaume:2016vff}:
for large \(Q\), the free energy scales like \(Q^{3/2}\), and we have an expansion in \(1/Q\).
The dependence on the manifold enters via the volume and the scalar curvature, with the latter controlling the first correction to the leading scaling. In the effective field theory, this dependence on the manifold is a consequence of the homogeneity of the ground state. However, within the \ac{eft} picture, the model-dependent coefficients are not accessible.  
In the path-integral formulation we adopted here, we can compute them from first principles.

\subsection{Finite charge, finite temperature: the unbroken phase}%
\label{sec:large-Q-finite-T}

Let us now consider the case of \(Q \neq 0\), \(T \neq 0\).
The crucial observation is that for finite \(\beta\) the zeta function \(\zeta(1| \theta, \Sigma, m)\) diverges at the massless point \(\theta = i m \beta\) (this is manifest in the expression in Eq.~\eqref{eq:zeta-1-finite-T}).
It follows that the zero modes \(A_i\) must vanish and the saddle point equations are
\begin{equation}
  \begin{dcases}
    \sum_{i=1}^N \zeta(1| \theta_i, \Sigma, m ) = 0, \\
    i Q_i + \frac{1}{s} \eval{ \pdv{\zeta(s| \theta_i, \Sigma, m)}{\theta_i}}_{s= 0} = 0, & i = 1, \dots, N.
  \end{dcases}
\end{equation}
We can think of this as a phase transition from the broken phase at \(T = 0\) to an unbroken phase where we still have \(U(N)\) symmetry.
Consistently with Goldstone's theorem, there are no massless modes and all the charge lives in the fluctuations (second equation). 
The transition happens at \(T = 0\) because we have chosen from the beginning to consider the \ac{ir} limit \(u \to \infty\) and there are no other scales.
From this point of view, we can think of this regime as describing a \ac{cft} at finite temperature.

For ease of notation we will consider in the following the case when there is a single fixed charge that rotates all the fields in the same way, so that all the \(Q_i\) and all the \(\theta_i\) are equal.
The saddle point equations~(\ref{eq:saddle}) reduce to
\begin{equation}
  \begin{cases}
    N \zeta(1| \theta, \Sigma, m ) = 0, \\
   i Q + N \frac{1}{s} \eval{ \pdv{\zeta(s| \theta, \Sigma, m)}{\theta}}_{s= 0} = 0.
  \end{cases}
\end{equation}
For simplicity we will also limit ourselves to the case of \(\Sigma = T^2\) so that the zeta function on \(S^1_\beta \times T^2\) can be written explicitly in terms of special functions as given in Eq.~(\ref{eq:zeta-T2-special-functions}):
\begin{equation}
    \zeta(s | \theta, T^2, m) = \frac{V \beta m^{3-2s}}{8 \pi^{3/2} \Gamma(s)}  \pqty{ \Gamma(s- \tfrac{3}{2}) + 4 \pqty{\frac{2}{m \beta}}^{3/2-s} \sum_{p=1}^\infty \frac{K_{s-3/2}(m \beta p)}{p^{3/2-s}}  \cos(p \theta)}.
\end{equation}
The saddle point equations become:\footnote{\(\Li_s(z)\) is the polylogarithm function \(\Li_s(z) = \sum_{k=1}^\infty \frac{z^k}{k^s} \). For \(s=1\), \(\Li_1(z) = - \log(1-z)\).}
\begin{equation}
  \begin{cases}
    m \beta + \log( \pqty{ 1 - e^{-m \beta - i \theta}} \pqty{1 - e^{-m \beta + i \theta}}) = 0, \\
    \frac{4 \pi}{V } Q + \frac{2  m N}{\beta} \log( \frac{1 - e^{-m \beta + i \theta}}{1 - e^{-m \beta - i \theta}} ) + \frac{2  N}{\beta^2} \pqty{ \Li_2(e^{-m \beta + i \theta}) - \Li_2(e^{-m \beta - i \theta})} = 0,
  \end{cases}
\end{equation}
and the expected \(2\pi\)-periodicity in \(\theta\) is manifest.
It is convenient to introduce the new variables
\begin{align}
  x &= e^{- m \beta}, & y &= e^{i \theta}
\end{align}
so that the first saddle point equation reduces to
\begin{align}
  x + \frac{1}{x} - y - \frac{1}{y} &= 1, & \text{\emph{i.e.} }  \cosh(m \beta) - \cos(\theta) &= \frac{1}{2} ,
\end{align}
where we see explicitly that \(m^2 + \theta^2/\beta^2 \neq 0\) so that there are no massless modes among the original fields \(\varphi_i\) and all the zero modes vanish \(A_i = 0\).

We are interested in the low-temperature (\( \beta \gg 1  \)) expansion, so it is convenient to solve the equation for \(y\). 
At low temperatures (\(x \to 0\)) this gives
\begin{align}
  y &= x + x^2+ \dots & \text{\emph{i.e.} }\theta &= i \pqty{ m \beta - e^{- m \beta} } + \dots.
\end{align}
This is encouraging since in the limit of zero temperature we recover the result \(\theta = i m \beta\) discussed in the previous section.
This behavior is consistent with a continuous phase transition at \(\beta = \infty\).

The other saddle point equation is transcendental, but can be solved perturbatively around \(\beta \to \infty\).
Using the expression of \(y = y(x)\) and expanding at first order in \(e^{- m \beta}\) we have (at leading order in \(N\))
\begin{equation}
  - \frac{4 \pi}{V} \frac{Q}{N}  + 2  m^2  + \frac{1}{3 \beta^2} \pi^2  - 2 \frac{e^{-m \beta}}{\beta}  \pqty{ m  + \frac{1 }{\beta} } + \order{e^{-2 m \beta}} = 0.
\end{equation}
Looking for a solution of the form
\begin{equation}
  m = m_0 + e^{-m_0 \beta} m_1
\end{equation}
we find
\begin{align}
  m_0^2 &=  \frac{4 \pi}{V} \frac{Q}{2N} - \frac{\pi^2}{6 \beta^2} , & m_1 &= \frac{1 }{2 m_0 \beta^2}  .
\end{align}
At leading order, neglecting the fluctuations \(\hat \lambda\), the free energy is
\begin{multline}
  F^{\saddle}(Q,\beta) = - \frac{V}{6 \pi} m^2 N  - i \frac{Q}{\beta} \theta - \frac{V N}{2 \pi \beta^3} \Big[ m \beta \pqty{ \Li_2(e^{-m \beta - i \theta}) + \Li_2(e^{-m \beta + i \theta})} \\
  + \pqty{ \Li_3(e^{-m \beta - i \theta}) + \Li_3(e^{-m \beta + i \theta})}\Big]   .
\end{multline}
Substituting the values of \(\theta\) and \(m\) at the fixed point and expanding in inverse powers of \(Q\) we find
\begin{multline}
  \frac{F^{\saddle}}{2N} = \frac{2}{3} \sqrt{\frac{4 \pi}{V} } \pqty{\frac{Q}{2N} }^{3/2} - \frac{\pi^2}{6 \beta^2} \sqrt{\frac{V}{4 \pi} } \pqty{\frac{Q}{2N} }^{1/2} - \frac{1}{\beta^3} \frac{V}{4 \pi} \zeta(3) + \order{Q^{-1/2}} \\
  +  \exp[ - \beta \pqty{\frac{Q}{2N} }^{1/2} \sqrt{\frac{4 \pi}{V} }] \pqty{\frac{1}{\beta^2} \sqrt{\frac{V}{4 \pi} }  \pqty{\frac{Q}{2N} }^{1/2} + \order{Q^0}} .
\end{multline}
In the limit \(\beta \to \infty \) we recover the broken-phase result of  Eq.~\eqref{eq:torus-energy-zero-T}.
The transition is continuous, even though in the broken phase all the charge is in the zero modes  while here it is in the fluctuations.

It is convenient to introduce a new parameter \(\rho\) as the total charge density divided by the rank of the symmetry group,
\begin{equation}
  \rho = \frac{Q}{2N} \frac{4 \pi}{V},  
\end{equation}
and the free energy density divided by the rank takes the simpler form
\begin{equation}
  \frac{4 \pi}{V} \frac{F^{\saddle}}{2N} = \frac{2}{3} \rho^{3/2} - \frac{\pi^2}{6 \beta^2} \rho^{1/2} - \frac{\zeta(3)}{\beta^3} + \dots + e^{- \beta \sqrt{\rho}} \pqty{ \frac{\rho^{1/2}}{\beta^2}  + \dots } .
\end{equation}
This is an expansion of the type \(F(\rho, \beta) = \rho^{3/2} f(1/ (\beta \rho^{1/2}))\), which is what one expects in general for a quantity of dimension \(3\) (the energy density) in a problem with two typical scales \(\rho\) and \(\beta\) in the limit where one of them (\(1/\beta\) in this case) is treated perturbatively.

Having the low-temperature expansion of the free energy, we can compute the corresponding entropy:
\begin{equation}
  \mathcal{S}^{\saddle} = \beta^2 \pdv{F}{\beta} = \frac{\pi}{6 } N V \frac{\rho^{1/2}}{\beta}  + \frac{3 N V \zeta(3)}{2 \pi \beta^2} + \dots - \frac{N V}{2 \pi} e^{- \beta \sqrt{\rho}} \pqty{ \rho + 2 \frac{\rho^{1/2}}{\beta} + \dots} .
\end{equation}
At zero temperature \(\beta \to \infty\), the entropy vanishes and this is consistent with the \ac{eft} result that the low-energy dynamics is controlled by an isolated ground state.

\bigskip

Finally, we can express the masses of the modes in Eq.~\eqref{eq:non-goldstone-dispersions} in terms of \(m\) and \(\beta\).
We find
\begin{align}
  M_+ &= 2 m + \dots  = 2 \rho^{1/2} + \dots  ; & M_- &= \frac{e^{- m \beta}}{\beta} + \dots = \frac{e^{-\beta \sqrt{\rho}}}{\beta} + \dots .
\end{align}
One of the masses is associated to the fixed charge, \(M_+ = \order{m} = \order{Q^{1/2}}\), the other one to the inverse temperature \(\beta\), \(M_- = \order{\tfrac{1}{\beta} e^{-\beta \sqrt{Q}}}\).
In the limit \(\beta \to \infty\), the latter becomes massless.
These are the Goldstone modes that we will discuss in detail in the next section.

\section{The Goldstones}%
\label{sec:goldstones}

Up to this point, \(N\) has been a generic parameter.
We have derived the saddle point equations and then computed the free energy assuming that the fluctuations \(\hat \lambda\) could be neglected.
This can be made more precise starting from the action~\eqref{eq:expanded-tr-log-original}.
In the standard treatment of the large-$N$ limit~\cite{PhysRevD.10.2491,ZinnJustin:2007zz}, a natural rescaling of the quantum fluctuations is introduced that results in a self-consistent $1/N$ expansion. In our case, we rescale the fluctuations as \(\hat \lambda \to \hat \lambda/N^{1/2}\).
In this way we introduce a hierarchy among the terms in the effective action. The results of the previous section are now understood as the leading effects in \(1/N\) and we can study the system perturbatively.
From now on we will take the limit
\begin{equation}
  N \gg 1 .
\end{equation}
This argument requires all the leading order terms in the action (which contribute to the saddle) to be of the same order \(\order{N}\).
In our case, this includes also the charge-fixing term \(i \theta Q\), which means that also \(Q\) is of order \(\order{N}\). In other words, we work at fixed $Q/N$.
This is in fact quite natural from our point of view:  \(Q\) is the sum of the \(N\) charges \(Q_i\) that we have fixed in the beginning and we take each of them to be of order \(\order{N^0}\).\footnote{This is to be contrasted with the situation in~\cite{Orlando:2019hte} where the fixed charge corresponds to an adjoint action and \(Q \) is of order \(\order{N^2}\).}

Our saddle-point result in Eq.~\eqref{eq:result-finite-Q-zero-T-zeta} is valid for any value of \(Q/N\). We have however observed that the zeta functions have a natural expansion in terms of the parameter \(Q/N \gg 1\).
This means that we are effectively working in the hierarchy
\begin{equation}
  1 \ll N \ll Q \ll N^2.
\end{equation}
Equivalently, we have two large numbers, \(N\) and \(Q/N\), with \(N \gg Q/N\) controlling the splitting between tree-level and quantum effects in the theory and \(Q/N\) giving an expansion of the physical observables at each fixed order in \(N\).

\bigskip

Of the phases discussed above, the broken phase is the most interesting because we expect new physics at the next-to-leading order.
One of the predictions of~\cite{Hellerman:2015nra,Alvarez-Gaume:2016vff} is that in the large-charge expansion of the free energy in this phase there is a universal \(Q^0\) term.
Its physical origin is the Casimir energy of a Goldstone mode with dispersion relation \(\omega = p/ \sqrt{2} \).
This dispersion relation is dictated by the tracelessness of the energy-momentum tensor, and we refer to this mode as the \emph{conformal Goldstone}.
Since this term is \(N\)-independent, it has to appear as part of the first \(1/N\) correction to the (order \(N\)) results of the previous section.
In the following we will consider the broken phase \(Q >0\), \(T = 0\) and 
will compute the \(\order{N^0}\) corrections pertinent to the Casimir energy.

\bigskip

In order to see the Goldstone modes explicitly we need to take a step back.
Consider the action for the \(u_i\) and use the fact that, as we have found in Section~\ref{sec:zero-Q-zero-T}, at the saddle point all the thetas are equal, \(\theta_i = i m \beta\):
\begin{equation}
  \begin{aligned}
    S_Q ={}& \eval{-i \theta Q + \pqty{\frac{\theta^2}{\beta^2} + m^2 } \frac{v^2}{2}V\beta}_{\theta = i m \beta } \\
    &+ \sum_{i=1}^N \int \dd{t}\dd{\Sigma} \bqty{ \pqty{ D_\mu u_i}^* \pqty{D_\mu u_i} + m^2 \abs{u_i}^2 + \frac{A_i}{\sqrt{2}} \hat \lambda \pqty{u_i + u_i^*} + \hat \lambda \abs{u_i}^2},
  \end{aligned}
\end{equation}
where \( Q \) is again the total charge and \(v^2\) is the sum of the squares of the \ac{vev},
\begin{align}
  Q &= \sum_{i=1}^N Q_i,  & v^2 = \sum_{i=1}^N A_i^2 .
\end{align}

It is convenient to decouple the mode which lies in the direction of the \ac{vev} \(A_i\). 
For this reason we introduce an orthonormal basis of \(\setC^N\) with generators that satisfy
\begin{align}
  e^0 &= \frac{(A_1, \dots, A_N)}{v}, & e^I \cdot e^0 &= 0, & e^I \cdot e^J &= \delta^{IJ}, & I, J = 1, \dots , N-1,
\end{align}
and we project the fields \(u^i\) on it. The Hamiltonian then reads
\begin{equation}
  \begin{aligned}
    S_Q ={}& m \beta Q + \int \dd{t} \dd{\Sigma} \bqty{\pqty{D_\mu u^0}^* \pqty{D_\mu u^0} + m^2 \abs{u^0}^2  + \frac{v}{\sqrt{2}} \hat \lambda \pqty{u^0 + (u^0)^*} + \hat \lambda \abs{u^0}^2} \\
    &+ \sum_{I = 1}^{N-1} \int \dd{t} \dd{\Sigma} \bqty{ \pqty{D_\mu u^I}^* \pqty{D_\mu u^I} + m^2 \abs{u^I}^2 + \hat \lambda \abs{u^I}^2 }.
\end{aligned}
\end{equation}
The \ac{vev} breaks the initial \(U(N)\) symmetry to the \(U(N-1)\) that rotates the fields \(u^I\).
This singles out naturally the field \(u^0\) which, together with the former Lagrange multiplier \(\hat \lambda\), will give rise to the expected conformal Goldstone mode.

\paragraph{Type-II Goldstones.}

The action for each of the $u^I$ is given by
\begin{equation}
  \label{eq:uI-action}
  S_Q[u_i] = \eval{ \int \dd{t} \dd{\Sigma} \bqty{\pqty{D_\mu u^I}^* \pqty{D_\mu u} + m^2\abs{u^I}^2+\hat \lambda \abs{u^I}^2}%
  }_{\theta = i m \beta} .
\end{equation}
This is expressed in terms of \(m^2\), which at the saddle point is proportional to our fixed control parameter \(Q/N\) (see Eq.~\eqref{eq:free-energy-zero-T-any-sigma}).
The inverse propagator reads
\begin{equation}
  \Delta^{-1} = \frac{1}{2} \begin{pmatrix}
    0                         & \omega^2 + p^2+ 2 i m \omega \\
    \omega^2+p^2 - 2i m \omega & 0
  \end{pmatrix}.
\end{equation}
We have two modes and the corresponding dispersion relations are given by the poles of the propagator:
\begin{align}
  \omega^2+\frac{p^4}{4m^2}+\dots &=0, & \omega^2+ 4 m^2 + 2p^2+\dots&=0.
\end{align}
The first mode is massless and describes a particle with quadratic dispersion relation.
We recognize the expected \((N-1)\) non-relativistic or type-II Goldstone modes of~\cite{Alvarez-Gaume:2016vff}.
These modes are naturally arranged into the fundamental representation of the unbroken \(U(N-1)\) that acts on the subset of generators \(e^I\), \(I=1,\dots, N-1\) of \(\setC^N\).
Each of these modes counts as two \ac{dof}~\cite{Nielsen:1975hm} as each of them is associated to a pair of unbroken Chevalley generators. In the charged background, these generators become canonically conjugate, \emph{i.e.} form a Heisenberg algebra, and result in only one field.
This result is valid for any value of the mass parameter \(m\), which itself does not need to be large.

\paragraph{The conformal Goldstone.}

The conformal Goldstone mode appears as a combination of \(u^0\) (from now on, for ease of notation \(u^0 = u\)) and \(\hat \lambda\).
Our strategy is to integrate out the other \(u^I\) and  write an effective action for these fields. We start again with the partition function
\begin{equation}
   Z(Q) = \int \DD{u} \DD{\lambda} \DD{u^I} e^{-S_Q[ u, \hat \lambda, u^I]} =  \int \DD{u} \DD{\lambda} e^{-S_Q[u, \hat \lambda]} ,
\end{equation}
where
\begin{equation}
  \label{eq:broken-phase-effective-action}
  \begin{aligned}
    S_Q[u, \lambda] ={}& m \beta  Q  + \pqty{N-1} \Tr[ \log(-D_\mu D^\mu + m^2)] \\
    &+ \int \dd{t} \dd{\Sigma} \Bigg[ \pqty{D_\mu u}^* \pqty{D_\mu u} + m^2 \abs{u}^2  + \frac{v}{\sqrt{2(N-1)}} \hat \lambda \pqty{u + u^*} \\
    & \hspace*{.25\textwidth} + \frac{ \hat \lambda}{\sqrt{N-1}} \abs{u}^2 \Bigg] 
    + \sum_{n=2}^\infty \frac{(-1)^{n+1}}{n (N-1)^{n/2-1} }  \Tr(\Delta \hat \lambda)^n.
  \end{aligned}
\end{equation}
We have decomposed \(\lambda\) as
\begin{equation}
  \lambda = m^2 - r + \frac{\hat \lambda}{\sqrt{N-1}}. 
\end{equation}

Being universal, the mode that we are looking for has to appear at order \(\order{N^0}\).
Assuming that all the charges that we have fixed at the beginning are of order one, \(Q_i = \order{N^0}\), we see that 
\(Q\) scales like \(\order{N}\).
Then the first line of Eq.~\eqref{eq:broken-phase-effective-action} is of order \(\order{N}\) (apart from the obvious \(-1\) piece), since at the saddle point the mass is of order \(m= \order{(Q/N)^{1/2}} = \order{N^0}\).
Moreover, the \ac{vev} is of order \(v =\order{(QN)^{1/4}} = \order{N^{1/2}}\), which means that the second line is of order \(\order{N^0}\).
The first one of the non-local terms is of order \(\order{N^0}\).
The other terms are suppressed at least as \(\order{N^{- 1/2}}\).
This is consistent with the general analysis at the beginning of this section.
The only limit that we need to take is \(N \gg 1\). The structure of the Goldstone fields remains the same for any value of  \(Q / N \).

\begin{figure}
  \centering
  \includegraphics{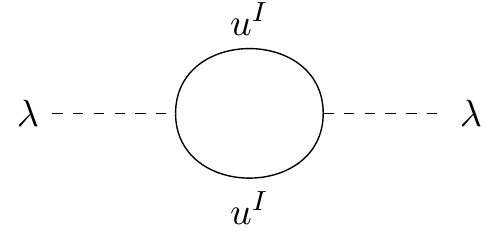}
  \caption{Feynman diagram representing the correction \(B(\omega, p)\) to the \(\hat \lambda\) propagator coming from a loop of the \(u^I\).}%
  \label{fig:bubble-diagram}
\end{figure}

  If we concentrate only on the terms that depend on \(u\) and \(\hat \lambda\), at order \(\order{N^0}\) we obtain the quadratic action 
\begin{equation}
  S^{(2)}[u, \hat \lambda] = \int \dd{t} \dd{\Sigma} \bqty{\pqty{D_\mu u}^* \pqty{D_\mu u} + m^2 \abs{u}^2  + \frac{v}{\sqrt{2 \pqty{N-1}}} \hat \lambda \pqty{u + u^*}} - \frac{1}{2} \Tr(\Delta \hat \lambda)^2 , 
\end{equation}
where \(m\) and \(v\) are the values at the saddle point.
It is convenient to write everything in terms of the fixed control parameter \(m\) using Eq.~\eqref{eq:saddle-finite-Q-no-T}:
\begin{equation}
  v^2 V  = -  \pqty{N-1} \zeta(\tfrac{1}{2}| \Sigma, m) ,
\end{equation}
and the Euclidean action is given by
\begin{multline}
  S^{(2)}[u, \hat \lambda] = \int \dd{t} \dd{\Sigma} \Bigg[ \del_\mu u^* \del_\mu u - m \pqty{ \del_0u^* u - u^* \del_0 u} \\
  + \sqrt{- \frac{\zeta(\tfrac{1}{2}| \Sigma, m)}{2V} } \hat \lambda \pqty{u + u^*} \Bigg] - \frac{1}{2} \Tr(\Delta \hat \lambda)^2 .
\end{multline} 
\(\Tr(\Delta\hat \lambda)^2\) is a non-local term that is completely expressed in terms of the propagator of the fields \(u^I\) (remember that this term comes from the expansion of the \(\Tr(\log(\cdot))\)):
\begin{equation}
  \Tr(\Delta \hat \lambda)^2 = \int \dd{t_1} \dd{\Sigma_1} \dd{t_2} \dd{\Sigma_2} \bqty{\hat \lambda(t_1, x_1) \hat \lambda(t_2, x_2) \Delta(t_1 - t_2, x_1 - x_2)^2 }.
\end{equation}
In terms of Feynman diagrams, this is the effect of a bubble of \(u^I\) on the propagator of \(\hat \lambda\) due to the interaction \(\hat \lambda \abs{u^I}^2\) in the action in Eq.\eqref{eq:uI-action}, see Figure~\ref{fig:bubble-diagram}.
In the \(u \to \infty\) limit that we have taken in order to be at the conformal point, this is the leading contribution to the propagator and its low energy limit is computed in Appendix~\ref{sec:one-loop-lambda-propagator}.
The resulting one-loop effective action at quadratic order is
\begin{multline}
  \label{eq:action-sigma-lambda-zero-T}
  S^{(2)}[u, \hat \lambda] = \int \dd{t} \dd{\Sigma}  \Bigg[ \del_\mu u^* \del^\mu u + m \pqty{u^* \del_0 u - u \del_0 u^*} + \sqrt{-\frac{\zeta(\tfrac{1}{2}|\Sigma,m)}{2 V} } \hat \lambda  \pqty{u + u^*} \\- \frac{\zeta(\tfrac{3}{2}|\Sigma,m)}{8V}\hat \lambda^2 \Bigg].
\end{multline}
There is no kinetic term for \(\hat \lambda\) so we can integrate it out:
\begin{equation}
  S^{(2)}[u] = \int \dd{t} \dd{\Sigma}  \bqty{\del_\mu u^* \del^\mu u + m \pqty{u^* \del_0 u - u \del_0 u^*} - \frac{\zeta(\tfrac{1}{2}|\Sigma,m)}{ \zeta(\tfrac{3}{2}|\Sigma,m)} \pqty{u + u^*}^2} .
\end{equation}
The leading-order contribution in the \(1/Q\) expansion of this term comes from the leading term in the heat kernel expansion of the zeta function,
\begin{equation}
  \zeta(s| \Sigma, m) = \frac{K_0}{s-1} m^{2-2s} + \dots
\end{equation}
and we find
\begin{equation}
  S^{(2)}[u ] = \int \dd{t} \dd{\Sigma}  \bqty{\del_\mu u^* \del^\mu u + m \pqty{u^* \del_0 u - u \del_0 u^*} + m^2 \pqty{u + u^*}^2}.
\end{equation}
The inverse propagator is
\begin{equation}
  \Delta^{-1} =
  \frac{1}{2} \begin{pmatrix}
    2 m^2 & \omega^2 + p^2 + 2 i m \omega + m^2 \\
    \omega^2 + p^2 - 2 i m \omega + m^2 & 2 m^2
  \end{pmatrix}.
\end{equation}
Its zeros describe a massless and a massive \ac{dof}, with dispersion relations
\begin{align}
  \omega^2 + \frac{1}{2}p^2 + \dots &= 0, & \omega^2 + 8 m^2 + \frac{3p^2}{2} + \dots &= 0.
\end{align}
The first one is the expected universal conformal Goldstone mode found in~\cite{Hellerman:2015nra}.
  In the limit where \(Q_i \gg 1\), it is controlled by a simple effective action:
\begin{align}
  S[\chi] = \int \dd{t} \dd{\Sigma} \bqty{(\del_t \chi)^2 + \tfrac{1}{2}(\grad\chi)^2} + \order{Q^{-1/2}} 
\end{align}
and its contribution to the free energy is
\begin{equation}%
  \label{eq:conformal-Goldstone-Casimir-energy}
  F[\chi] = \frac{1}{2 \sqrt{2}} \zeta(-\tfrac{1}{2}|\Sigma, 0),
\end{equation}
where the \(1/(2 \sqrt{2})\) is characteristic of a real scalar of velocity \(1/\sqrt{2}\).
In the case of the two-sphere, using  Eq.~\eqref{eq:zeta-S2-Carletti} we find \(F[\chi] = -0.0937\dots\).

\section{Conclusions}%
\label{sec:conclusions}

This article is dedicated to exploring the symmetry-breaking properties of the conformal $O(2N)$ vector model
 in the large-charge limit as presented in~\cite{Hellerman:2015nra,Alvarez-Gaume:2016vff}, but 
considering the dependence on the number of fields.  We started with a first-principles definition of the theory, and then proceeded to study the interplay between the
large-$N$ and large-$Q$ limits.  We recover previous results using different
techniques, which we expect to be able to use to systematically analyze many other theories, 
with and without supersymmetry
(for supersymmetric theories, see for instance the results in~\cite{Hellerman:2017sur,Hellerman:2018xpi,Grassi:2019txd}).  

Our approach allows us moreover to work at finite temperature.
We observe for $u=\infty$ that the symmetry breaking takes place at $T=0$, while at finite temperature there is an unbroken phase.
The phase transition between the two phases is continuous.

\bigskip

The non-trivial resummation of quantum effects of
the theory which takes place in our approach may be at the origin of a small puzzle: In~\cite{Alvarez-Gaume:2016vff} we found that in the simplest \ac{eft} the product of the two leading coefficients in the large-charge expansion is
$c_{3/2} c_{1/2} =1/12$; in this paper we find $c_{3/2} c_{1/2} =1/9.$
This difference may be due to a different splitting between tree-level and quantum corrections
in the two treatments.
In this paper however, we have used the large-$N$ expansion to
explore systematically a region of the original theory that may have
couplings $\sim \order{1}$, and  the standard large-$N$ lore would
indicate this computation to be more reliable. 

\bigskip

Starting from our results, there are a number of open questions to address in the future:
\begin{itemize}
\item We have compared our predictions for $c_{3/2}$ and $c_{1/2}$ to lattice data for $N=2$. Future lattice studies for $N>2$ will allow a more accurate comparison with our predictions.
\item In this article, we have computed the leading order in $N$ and the contributions of the Goldstones to the subleading corrections. It would be interesting to calculate further subleading corrections in the $1/N$ expansion, which might also shed light on the puzzle regarding the product $c_{3/2} c_{1/2}$.
\item We have concentrated on the limit \(Q/N \gg 1\), but our construction remains valid for any value of this ratio. In Eq.~\eqref{eq:conformal-dimension-small-charge} we have found a small-charge expansion for the conformal dimensions.
  This regime certainly deserves a more detailed study.
\item Recently, the $4-\epsilon$ expansion at large charge has been studied~\cite{Arias-Tamargo:2019xld,1752534,1752533}, making use of a double-scaling limit where $Q\epsilon$ was held fixed. It would  be interesting to explore $\epsilon$-expansions in the present context.
\item Here, we focused on the $O(2N)$ vector model. One could extend the large-$N$ treatment also to theories at large charge with matrix-valued fields, such as the ones studied in~\cite{Loukas:2017lof,Loukas:2017hic,Orlando:2019hte}.
\item The large-charge expansion can be used also to discuss dualities, that we typically expect to become perturbatively treatable in the large-charge limit.
  Interestingly, a construction similar to the one used in this paper with systems at large-\(N\) and fixed charge \(Q\) has appeared in the literature~\cite{Filothodoros:2016txa,Filothodoros:2018pdj} in connection with Fermi-boson duality.
  This seems to be a promising starting point for both a large-charge and an \ac{eft} analysis.
\item In this paper, we have worked directly at criticality.
  It is however possible to work at finite $\abs{\varphi}^4$ coupling $u$ and to follow the \ac{rg} flow. A similar double-scaling limit as alluded to above might be helpful for this and helpful in computing strong/weak ratios of observables such as the entropy, as discused for example in~\cite{Romatschke:2019ybu,DeWolfe:2019etx}.
\end{itemize}

\subsection*{Acknowledgments}

D.O. would like to thank Igor Pesando for useful discussions.
The work of S.R. is supported by the Swiss National Science Foundation under grant number \textsc{pp00p2\_183718/1}.
D.O. acknowledges partial support by the \textsc{nccr 51nf40--141869} ``The Mathematics of Physics'' (Swiss\textsc{map}).
D.O. and S.R. would like to thank the Simons Center for Geometry and Physics for  hospitality during the final phases of this work.

\newpage
\appendix
\section{The zeta function on \(S^1_\beta \times \Sigma \)}%
\label{sec:S1-Sigma-zeta-function}

In this appendix we discuss the zeta function for the operator \(-D_\mu D^\mu + m^2\), or equivalently the operator \(-\del_t^2 - \Laplacian + m^2\) on \(S_\beta^1 \times \Sigma\) with twisted boundary conditions \(\varphi( \beta, x) = e^{i \theta} \varphi(0, x)\) in the limit of \(m \gg 1\).

Let \(E(p)^2\) be the eigenvalues of \(\Sigma\) (that we assume compact).
Then the eigenvalues of our operator are
\begin{equation}
  \spec( -D_\mu D^\mu + m^2) = \set{\pqty{\frac{2 \pi n}{\beta} + \frac{\theta}{\beta} }^2 + E(p)^2 + m^2| n \in \setZ , p \in \spec(\Laplacian_\Sigma)}.
\end{equation}
Using the Mellin representation we can write:
\begin{equation}
  \zeta(s| \theta, \Sigma, m) = \frac{1}{\Gamma(s)} \int_0^\infty \frac{\dd{t}}{t} t^s e^{-m^2 t } \sum_{n \in \setZ} e^{- \pqty{\frac{2 \pi n}{\beta} + \frac{\theta}{\beta} }^2 t} \sum_p e^{-E(p)^2 t} .
\end{equation}
If \(m \gg 1\) the integral localizes around \(t =0\).
This allows us to decouple the \(S^1 \) contribution from the \(\Sigma \) contribution.

For the time part, we observe that this is a theta function, and we can Poisson resum it in order to obtain an expansion in \(e^{-1/t}\) that can be easily expanded for small \(t\):
\begin{equation}
  \label{eq:Poisson-resummation}
  \sum_{n \in Z} e^{-  \pqty{\frac{2 \pi n}{\beta} + \frac{\theta}{\beta} }^2 t} = \thetafunc{\frac{\theta}{2\pi}}{0}{0}{\tfrac{4 \pi i t}{\beta^2} } = \frac{1}{2 \sqrt{\pi}} \frac{\beta}{ t^{1/2}} \pqty{1 + 2 \sum_{p=1}^\infty \cos(p \theta) e^{- p^2 \beta^2/(4  t)} } .
\end{equation}

In this form, we see explicitly that the zeta function is $2\pi$-periodic in $\theta$ as we knew from the general arguments given in Section~\ref{sec:action}.

In the limit \(t \to 0 \) it is convenient to expand the \(\Sigma\)-dependent term using Weyl's asymptotic formula~\cite{weyl1911asymptotische,rosenberg_1997,Vassilevich:2003xt}
\begin{equation}
  \label{eq:Weyl-asymptotic-formula}
  \Tr(e^{\Laplacian_\Sigma{} t}) = \sum_{n=0}^{\infty} K_n t^{n/2 - 1},
\end{equation}
where \(K_n\) are the heat kernel coefficients.
These coefficients can be expressed in terms of the geometry of the manifold.
In general, if \(\Sigma\) has no boundary the odd coefficients vanish, \(K_{2n +1 } = 0\).
The important ones for our calculations are
\begin{align}
  K_0 &= \frac{1}{(4 \pi)^{d/2}} V , & K_2 &= \frac{1}{6 (4\pi)^{d/2}} \int R \dd{\Sigma}
\end{align}
where \(d\) is the number of dimensions (for us \(d =2\)), \(V\) is the volume and \(R\) is the scalar curvature of \(\Sigma\).

We can now write the large-\(m\) expansion of \(\zeta(s | \theta, \Sigma, m)\) in terms of the zeta function on the \(S^1\) alone:
\begin{equation}%
  \label{eq:zeta-on-theta-sigma}
\begin{aligned}
  \zeta(s | \theta, \Sigma, m) &= \frac{1}{\Gamma(s)} \int_0^\infty \frac{\dd{t}}{t} t^s e^{-m^2 t } \sum_{n \in \setZ} e^{- \pqty{\frac{2 \pi n}{\beta} + \frac{\theta}{\beta} }^2 t} \sum_{n=0}^{\infty} K_n t^{n/2 - 1} \\  
  &= \sum_{n=0}^{\infty} K_n \frac{\Gamma(s + \frac{n}{2} - 1)}{\Gamma(s)} \zeta(s + \tfrac{n}{2} - 1 | \theta, m).%
\end{aligned}
\end{equation}
The zeta function on the twisted circle can be evaluated explicitly:
\begin{equation}
  \begin{aligned}
    \zeta(s | \theta, m) &= \frac{1}{\Gamma(s)} \int_0^\infty \frac{\dd{t}}{t} t^s e^{-m^2 t} \theta \bqty{\genfrac{}{}{0pt}{1}{\theta/(2\pi)}{0}}\pqty{0, \frac{4 \pi i t}{\beta^2} } \\
    &= \frac{1}{\Gamma(s)} \int_0^\infty \frac{\dd{t}}{t} t^s e^{-m^2 t} \frac{1}{2 \sqrt{\pi}} \frac{\beta}{ t^{1/2}} \pqty{1 + 2 \sum_{p=1}^\infty \cos(p\theta) e^{- p^2 \beta^2/(4  t)} }  \\
     &= \frac{\beta m^{1-2s}}{2 \sqrt{\pi} \Gamma(s)} \pqty{ \Gamma(s- \tfrac{1}{2}) + 4 \pqty{\frac{2}{m \beta}}^{1/2-s} \sum_{p=1}^\infty \frac{K_{s-1/2}(m \beta p)}{p^{1/2-s}}  \cos(p \theta)},
  \end{aligned}
\end{equation}
where \(K_s(z)\) is the modified Bessel function of the second kind that for the values we are interested in takes the simple form
\begin{align}
  K_{\pm 1/2}(z) &= \sqrt{\frac{\pi}{2}} \frac{e^{-z}}{z^{1/2}}  , &
  K_{\pm 3/2}(z) &= \sqrt{\frac{\pi}{2}} \frac{e^{-z}}{z^{3/2}} ( 1 + z) .
\end{align}

If we just keep the first two terms \(K_0\) and \(K_2\) in the heat-kernel expansion we can evaluate explicitly the zeta function for the values that we need in the saddle equations and in the free energy:
\begin{align}
  \eval{  \dv{\zeta(s|\theta, \Sigma, m)}{s}}_{s=0} &=
                                        \begin{aligned}[t]
                                          K_0 \Big( \frac{2 m^2 \beta}{3} + 2 \frac{m}{\beta} \pqty{Li_2(e^{-m \beta + i \theta}) + Li_2(e^{-m \beta - i \theta})} \\
                                          + \frac{2}{\beta^2} \pqty{Li_3(e^{-m \beta + i \theta}) + Li_3(e^{-m \beta - i \theta})} \Big) \\
                                          -K_2 \log(2 (\cosh(m \beta) - \cos(\theta)))
                                        \end{aligned} \\
  \eval{ \frac{1}{s}  \dv{\zeta(s|\theta, \Sigma, m)}{\theta}}_{s=0} &=
                                                    \begin{aligned}[t]
                                                       \frac{2 i K_0}{\beta^2} \pqty{ m \beta \log(\frac{1 - e^{-m \beta - i \theta}}{1 - e^{-m \beta + i \theta}} ) + Li_2(e^{-m \beta + i \theta}) - Li_2(e^{-m \beta - i \theta})}  \\
                                                       - K_2 \frac{\sin(\theta)}{\cosh(m \beta) - \cos(\theta)}
                                                    \end{aligned} \\
  \label{eq:zeta-1-finite-T}
  \zeta(1| \theta, \Sigma, m) &= -K_0 \log(2 (\cosh(m \beta) - \cos(\theta))) + K_2 \frac{\beta \sinh(m \beta)}{2m(\cosh(m \beta) - \cos(\theta))} 
\end{align}

\paragraph{Zero temperature.}
In the limit \(\beta \to \infty\) we can approximate the theta function keeping only the first term:
\begin{equation}
  \thetafunc{\frac{\theta}{2\pi}}{0}{0}{\tfrac{4 \pi i t}{\beta^2} } \approx \frac{1}{2 \sqrt{\pi}} \frac{\beta}{ t^{1/2}}.
\end{equation}
This means that the zeta function on \(S^1_\infty \times \Sigma\) is rewritten in terms of the zeta function on \(\Sigma\) alone:
\begin{equation}
  \label{eq:zeta-for-beta-infinity}
  \begin{aligned}
    \zeta(s | \theta, \Sigma, m) &= \frac{1}{\Gamma(s)} \int_0^\infty \frac{\dd{t}}{t} t^s e^{-m^2 t } \sum_{n \in Z} e^{- \pqty{\frac{2 \pi n}{\beta} + \frac{\theta}{\beta} }^2 t} \sum_p e^{-E(p)^2 t}  \\
    &\approx \frac{1}{\Gamma(s)} \int_0^\infty \frac{\dd{t}}{t} t^s e^{-m^2 t }  \frac{1}{2 \sqrt{\pi}} \frac{\beta}{\ell t^{1/2}} \sum_p e^{-E(p)^2 t} + \dots  \\
    &= \beta \frac{\Gamma(s-\frac{1}{2})}{2 \sqrt{\pi} \Gamma(s)}  \zeta(s-\tfrac{1}{2}| \Sigma, m) . %
  \end{aligned}
\end{equation}

\section{The zeta function on the torus and the two-sphere}%
\label{sec:sphere-zeta-function}

In this appendix we want to write an asymptotic expression for the zeta function for the operator \(-\Laplacian + m^2\) on the two-torus and the two-sphere.
In general, given the eigenvalues \(E(p)^2\) of the Laplacian we can write the 
zeta function as a Mellin transform,
\begin{equation}
  \zeta(s | S^2, m) = \frac{1}{\Gamma(s)} \int \frac{\dd{t}}{t} t^s \Tr[ e^{ \pqty{\Laplacian - m^2}t} ] = \frac{1}{\Gamma(s)} \int \frac{\dd{t}}{t} t^s e^{-m^2 t } \sum_{p} e^{- E(p)^2 t} .
\end{equation}
We will be mainly interested in the limit of \(m \gg 1\), where we expect the integral to be localized around \(t = 0\).
In this region we can use Weyl's asymptotic formula in Eq.~\eqref{eq:Weyl-asymptotic-formula} to express the trace and find immediately that
\begin{equation}
  \label{eq:zeta-on-Sigma}
  \zeta(s | \Sigma, m) = \sum_{n=0}^\infty K_n \frac{\Gamma(s+\tfrac{n}{2}-1)}{\Gamma(s)} m^{2-n-2s} .
\end{equation}

\paragraph{The torus.}
When \(\Sigma = T^2\), the only non-vanishing heat kernel coefficient is \(K_0 = V/(4 \pi)\), so we obtain immediately%
\begin{equation}
  \label{eq:zeta-T2-zero-T}
  \zeta(s | T^2, m) = \frac{V}{4 \pi \pqty{s - 1}} m^{2-2s},
\end{equation}
and for \(S^1_\beta \times T^2\), using Eq.~\eqref{eq:zeta-on-theta-sigma}:
\begin{equation}
  \label{eq:zeta-T2-special-functions}
    \zeta(s | \theta, T^2, m) = \frac{V \beta m^{3-2s}}{8 \pi^{3/2} \Gamma(s)}  \pqty{ \Gamma(s- \tfrac{3}{2}) + 4 \pqty{\frac{2}{m \beta}}^{3/2-s} \sum_{p=1}^\infty \frac{K_{s-3/2}(m \beta p)}{p^{3/2-s}}  \cos(p \theta)}.
\end{equation}

\paragraph{The two-sphere.}

In the case of the two-sphere of unit radius \(S^2\) we need to distinguish two limits, \(m \to 0\) and \(m \to \infty\).
In either case we will obtain an expansion in \(m^2 - 1/4\) which can be understood in terms of conformal coupling: \(1/4\) is precisely \(R/8\) for a unit sphere, so \(m^2 - 1/4\) is the squared mass for a conformally-coupled scalar field (this is of course a manifestation of the Breitenlohner--Freedman bound~\cite{Breitenlohner:1982bm}).
 
\bigskip

In the \(m \to 0\) limit we can use a binomial expansion valid for \(0< m^2 < 1/2\).
\begin{multline}
  \label{eq:zeta-S2-small-m}
  \zeta(s | S^2, m) = \sum_{l = 0}^\infty \pqty{2l + 1} \pqty{ l ( l + 1 ) + m^2}^{-s} %
  = 2 \sum_{l=0}^\infty \pqty{l + \tfrac{1}{2} }^{-2s + 1} \pqty{1 + \frac{m^2 - \frac{1}{4} }{\pqty{l + \frac{1}{2} }^2}}^{-s} \\
  = 2 \sum_{l=0}^{\infty} \pqty{l + \tfrac{1}{2} }^{-2s+1} \sum_{k=0}^\infty \binom{-s}{k} \pqty{\frac{m^2 - \frac{1}{4} }{\pqty{l + \frac{1}{2} }^2}}^k \\
  =2 \sum_{k=0}^{\infty} \sum_{l=0}^{\infty} \binom{-s}{k} \pqty{l + \tfrac{1}{2} }^{-2s - 2k + 1} \pqty{m^2 - \tfrac{1}{4} }^{k} \\
  = 2 \sum_{k=0}^{\infty} \binom{-s}{k} \zeta(2s + 2k - 1, \tfrac{1}{2} )  \pqty{m^2 - \frac{1}{4} }^{k} .
\end{multline}
The result is a series in \(m^2 - 1/4\), with coefficients expressed in terms of Hurwitz zeta functions \(\zeta(s,a)\).
In the cases of interest, \(s = \pm 1/2\) and we can use the fact that
\begin{equation}
  \zeta(2n, \tfrac{1}{2} ) = \pqty{2^{2n} -1} \zeta(2n) = (-1)^{n+1} \pqty{2^{2n} -1} \frac{B_{2n}(2\pi)^{2n}}{2 (2n)!} ,
\end{equation}
where \(B_{2n}\) are the Bernoulli numbers.
The first terms of the series are then
\begin{align}
  \zeta(-\tfrac{1}{2} | S^2, m ) &= - \frac{\pi^2}{8} \pqty{m^2 - \frac{1}{4} }^{2} + \frac{\pi^4}{96} \pqty{m^2 - \frac{1}{4} }^{3} + \dots  \\
  \zeta(\tfrac{1}{2} | S^2, m ) &= - \frac{\pi^2}{2} \pqty{m^2 - \frac{1}{4} } + \frac{\pi^4}{8} \pqty{m^2 - \frac{1}{4} }^{2} + \dots 
\end{align}
In the special case of \(m = 0\), one needs to separate the mode \(l = 0\) and finds
\begin{equation}
  \label{eq:zeta-S2-Carletti}
  \begin{aligned}
  \zeta(s | S^2, 0) &= 2 \sum_{k=0}^{\infty} \sum_{l=1}^{\infty} \binom{-s}{k} \pqty{l + \frac{1}{2} }^{-2s - 2k + 1} \pqty{- \frac{1}{4} }^{k} \\
  &= \sum_{k=0}^{\infty} (-1)^k \binom{-s}{k}  2^{1-2k} \zeta(2s +2k -1, \tfrac{3}{2})
\end{aligned}
\end{equation}
This is the expression obtained in~\cite{10.2307/2161165}.
The series converges very rapidly and can be evaluated numerically to give \(1/(2 \sqrt{2}) \zeta(-\tfrac{1}{2}|S^2,0)= -0.0937254\dots \) for the zero-point energy of the conformal Goldstone in Eq.~\eqref{eq:conformal-Goldstone-Casimir-energy}.

\bigskip

\begin{figure}
  \centering
  \begin{tikzpicture}
    \node at (-.25\textwidth, 0) {\includegraphics[width=.45\textwidth]{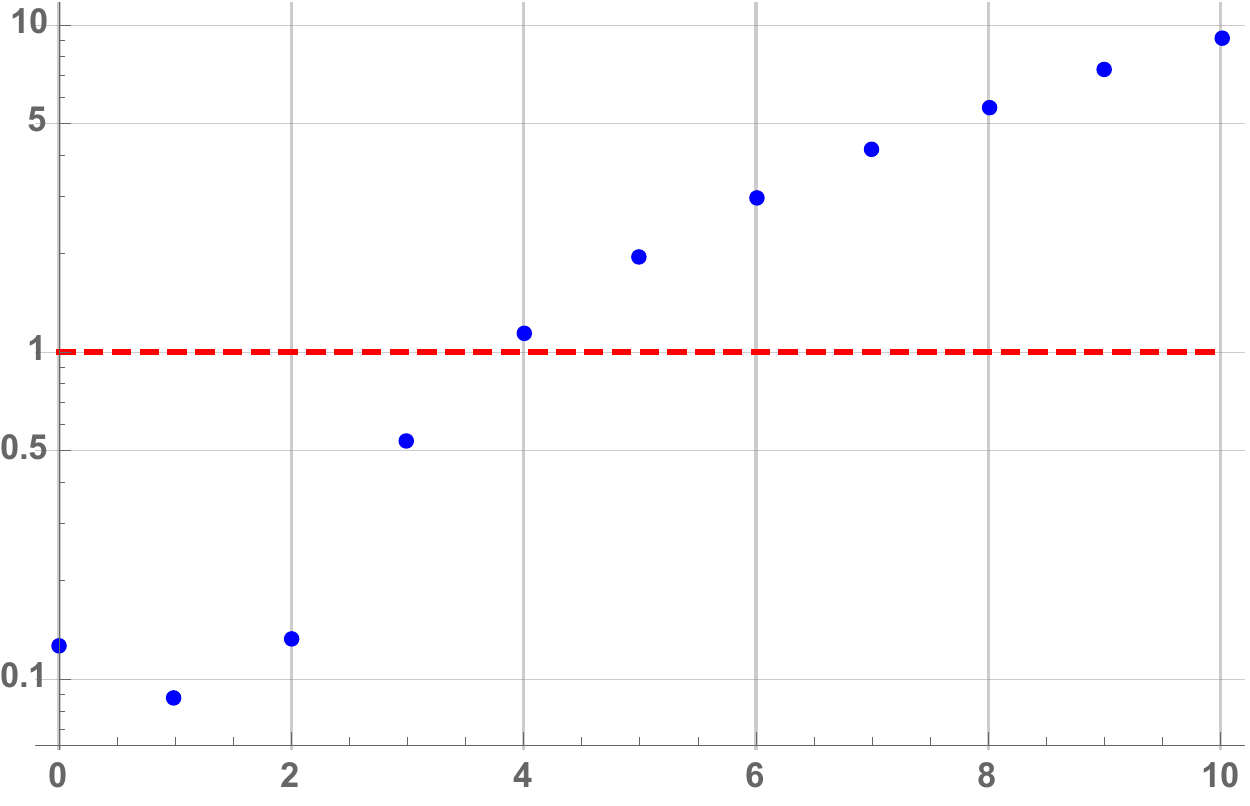}};
    \node at (.25\textwidth, 0) {\includegraphics[width=.45\textwidth]{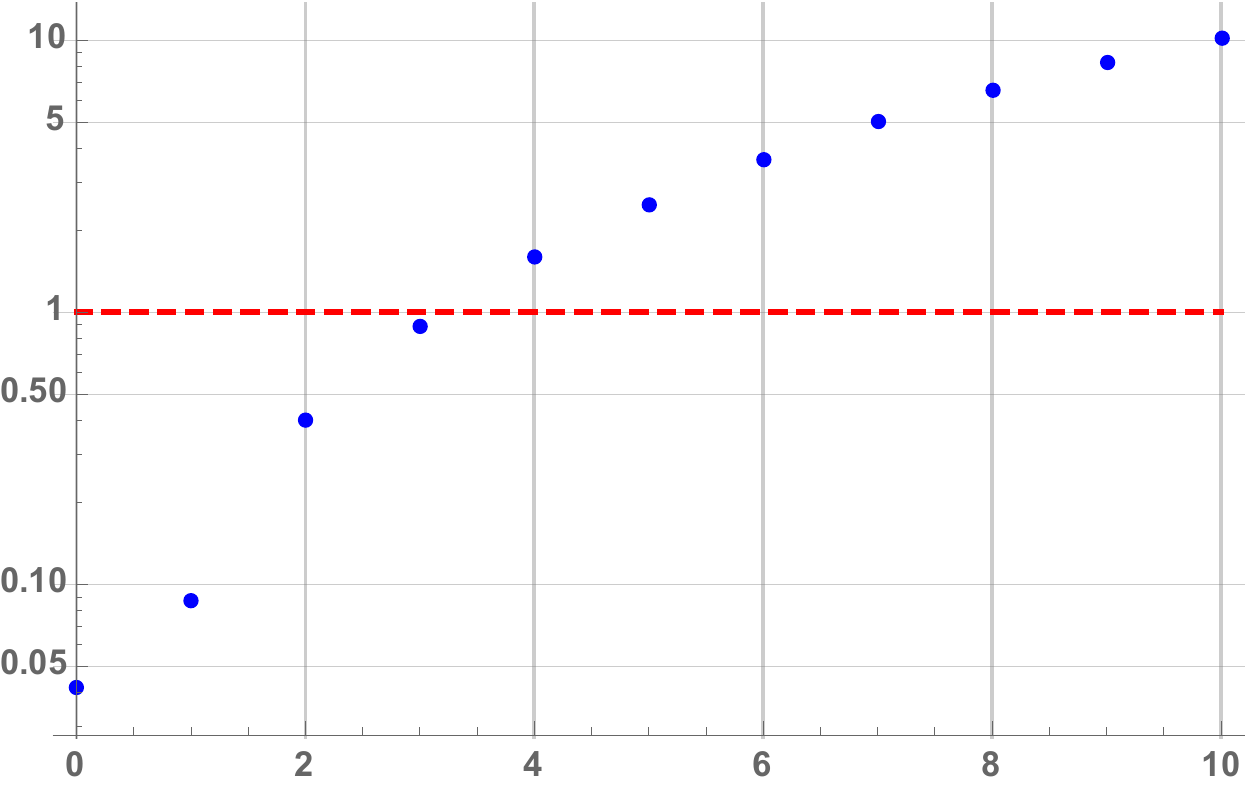}};

    \begin{small} \sffamily
    \node at (-.25\textwidth, -2.75) {(a) \(s = - 1/2\)};
    \node at (.25\textwidth, -2.75) {(b) \(s =  1/2\)};
  \end{small}
  \end{tikzpicture}
  \caption{Ratio of two consecutive coefficients in the asymptotic expansion of \(\zeta(s| S^2)\) for \(s = \pm 1/2\). In both cases the optimal truncation is at  the fourth term. After that coefficient of the next-order term becomes larger than the previous one.}%
  \label{fig:zeta-function-coefficient-ratios}
\end{figure}

In the opposite limit of \(m \to \infty\), we start from the asymptotic expression for the trace of the exponential~\cite{McKean:1967xf}:
\begin{equation}
  \Tr[e^{\Laplacian t}] = \sum_{\ell = 0}^\infty \pqty{2 \ell + 1} e^{-\ell \pqty{\ell+1} t} = 2 \frac{e^{t/4}}{\sqrt{\pi} t^{3/2}} \int_0^1 \dd{y} y \frac{e^{-y^2/t}}{\sin(y)}.
\end{equation}
We can write this integral in terms of an asymptotic expansion for small \(t\),
\begin{equation}
  2 \int_0^1 \dd{y} y \frac{e^{-y^2/t}}{\sin(y)} = \int_{-1}^1 \dd{y} y \frac{e^{-y^2/t}}{\sin(y)} \approx \int_{-\infty}^{\infty} \dd{y} y \frac{e^{-y^2/t}}{\sin(y)} + R(t),
\end{equation}
where \(R(t )\) is an exponential correction. 
The integral can be computed expanding the integrand in series,
\begin{equation}
  \frac{y}{\sin(y)} = \sum_{n=0}^\infty \frac{(-1)^{n+1 } 2\pqty{2^{2n-1} - 1}B_{2n}}{(2n)!} y^{2n}.
\end{equation}
We can rewrite the zeta function as
\begin{multline}
  \zeta(s | S^2, m) = \frac{1}{\Gamma(s)} \int_0^\infty \frac{\dd{t}}{t} t^s e^{-m^2 t} \Tr[ e^{\Laplacian t} ] \\
  = \frac{1}{\Gamma(s)} \int_0^\infty \frac{\dd{t}}{t} t^s e^{-m^2 t} \frac{e^{t/4}}{\sqrt{\pi} t^{3/2}} \int_{-\infty}^{\infty} \dd{y} e^{-y^2/t} \sum_{n=0}^\infty \frac{(-1)^{n+1 } 2\pqty{2^{2n-1} - 1}B_{2n}}{(2n)!} y^{2n}.
\end{multline}
Exchanging the order of sum and integration we can solve the integrals:
\begin{multline}
  \label{eq:zeta-S2-integral-representation}
  \zeta(s | S^2, m) = \frac{1}{\Gamma(s)} \int_0^\infty \frac{\dd{t}}{t} t^s e^{-m^2 t} \frac{e^{t/4}}{\sqrt{\pi} t^{3/2}} \int_{-\infty}^{\infty} e^{-y^2/t} \sum_{n=0}^\infty \frac{(-1)^{n+1 } 2\pqty{2^{2n-1} - 1}B_{2n}}{(2n)!} y^{2n} \\
  = \sum_{n=0}^\infty \frac{(-1)^{n+1 } 2\pqty{2^{2n-1} - 1}B_{2n}}{(2n)!}  \frac{1}{\Gamma(s)} \int_0^\infty \frac{\dd{t}}{t} t^s e^{-m^2 t} \frac{e^{t/4}}{\sqrt{\pi} t^{3/2}} \int_{-\infty}^{\infty} e^{-y^2/t} y^{2n}.
\end{multline}
The integral over \(y\) is Gaussian, and the integral over \(t\) is expressed in terms of gamma functions.
The final result is
\begin{equation}
  \label{eq:zeta-S2-zero-T}
  \begin{aligned}
    \zeta(s | S^2, m) %
    & = \pqty{m^2 - \frac{1}{4}}^{1-s} \sum_{n=0}^\infty \frac{(-1)^{n+1} \pqty{1 - 2^{1 - 2n}} B_{2n}}{ n + s - 1} \binom{n + s -1}{n}\frac{1}{\pqty{m^2 - \frac{1}{4}}^n} \\
    &= \frac{1}{s - 1} \pqty{m^2 - \frac{1}{4}}^{1-s} + \frac{1}{12} \pqty{m^2 - \frac{1}{4}}^{-s} + \frac{7 s}{480} \pqty{m^2 - \frac{1}{4}}^{-1-s} + \dots
  \end{aligned}
\end{equation}
The expansion is asymptotic and the optimal truncation depends on \(s\).
In the cases of interest, \(s = \pm 1/2\), the optimal truncation is at the fourth term (see Figure~\ref{fig:zeta-function-coefficient-ratios}).

Finally, we can derive an asymptotic expansion for the heat kernel coefficients:
\begin{equation}
  \label{eq:heat-kernel-sphere}
  \begin{aligned}
    \Tr[e^{\Laplacian t}] &=  2 \frac{e^{t/4}}{\sqrt{\pi} t^{3/2}} \int_0^1 \dd{y} y \frac{e^{-y^2/t}}{\sin(y)} = \sum_{n=0}^\infty \pqty{ \sum_{p=0}^n \frac{(-1)^{p+1} \pqty{-2+4^p}B_{2p}}{\Gamma(p+1) \Gamma(1+n-p)}} \frac{1}{4^n} t^{n-1} \\
    &= \frac{1}{t} + \frac{1}{3} + \frac{1}{15} t + \frac{4}{315} t^2 + \dots
  \end{aligned}
\end{equation}
and use it to write the zeta function \(\zeta(s|\theta, S^2, m)\) using  Eq.~\eqref{eq:zeta-on-theta-sigma}.

\section{The one-loop term in the propagator for \(\hat \lambda\)}%
\label{sec:one-loop-lambda-propagator}

In this appendix we compute the low-energy contribution to the \(\lambda\) propagator coming from a loop of \(u^i\) (see Figure~\ref{fig:bubble-diagram}).
\begin{equation}
  \Tr(\Delta \hat \lambda)^2 = \int \dd{t_1} \dd{\Sigma_1} \dd{t_2} \dd{\Sigma_2} \hat \lambda(t_1, x_1) \hat \lambda(t_2, x_2) \Delta(t_1 - t_2, x_1 - x_2)^2 .
\end{equation}

As usual, it is convenient to go to Fourier space and rewrite the square of the propagator as
\begin{equation}
  \Delta(t - t', x - x')^2 = \int \frac{\dd{\omega}}{2 \pi} \sum_p e^{i \omega(t-t')} \phi_p(x - x') B(\omega, p),
\end{equation}
where \(\set{\phi_p}\) is a complete set of eigenfunctions of the Laplacian \(\Laplacian_\Sigma\).
For a generic manifold, the expression for \(B(\omega,p)\) is quite involved since it depends on the product of three functions \(\phi_p(x)\) (on the sphere, for example, it depends on Wigner's \(3j\) symbol).
In order to recover the Goldstone mode, we concentrate on the low-energy behavior and  just need to compute \(B(0,0)\):
\begin{equation}
  B(0,0) = \frac{\beta}{2\pi} \int \dd{\omega} \sum_p \Delta(\omega,p) \Delta(-\omega, -p) .
\end{equation}
By definition the propagator \(\Delta(\omega, p)\) is
\begin{equation}
  \Delta(\omega, p) = \frac{1}{\pqty{\omega + \frac{\theta}{\beta}}^2 + E(p)^2 + m^2} ,
\end{equation}
In the zero-temperature limit we can apply our zeta-function regularization scheme (the computation is similar to the one in Eq.~\eqref{eq:zeta-beta-infinity}):
\begin{equation}
  B(0,0) = \frac{\beta}{2\pi}  \int \dd{\omega} \sum_p \frac{1}{\pqty{\omega^2 + E(p)^2 + m^2}^2}  = \frac{\beta}{4} \sum_p \frac{1}{\pqty{E(p)^2 + m^2}^{3/2}} = \frac{\beta}{4} \zeta(\tfrac{3}{2}| \Sigma, m).
\end{equation}
Putting this back in the expression for \(\Tr(\Delta \hat \lambda)^2\) and using the orthogonality of the basis of the \(\phi_p\) we find
\begin{equation}
  \Tr( \Delta \hat \lambda)^2 \approx \int \frac{\dd{\omega}}{2\pi}  \sum_p \hat \lambda(\omega, p) \hat \lambda(-\omega, -p) \frac{\beta}{4V } \zeta(\tfrac{3}{2}| \Sigma, m).
\end{equation}

\setstretch{1}

\printbibliography{}

@article{Nicolis:2013sga,
	Archiveprefix = {arXiv},
	Author = {Nicolis, Alberto and Penco, Riccardo and Piazza, Federico and Rosen, Rachel A.},
	Doi = {10.1007/JHEP11(2013)055},
	Eprint = {1306.1240},
	Journal = {JHEP},
	Pages = {055},
	Primaryclass = {hep-th},
	Slaccitation = {%%CITATION = ARXIV:1306.1240;%%},
	Title = {{More on gapped Goldstones at finite density: More gapped Goldstones}},
	Volume = {11},
	Year = {2013}}

@article{Nicolis:2013lma,
	Archiveprefix = {arXiv},
	Author = {Nicolis, Alberto and Penco, Riccardo and Rosen, Rachel A.},
	Doi = {10.1103/PhysRevD.89.045002},
	Eprint = {1307.0517},
	Journal = {Phys. Rev.},
	Number = {4},
	Pages = {045002},
	Primaryclass = {hep-th},
	Slaccitation = {%%CITATION = ARXIV:1307.0517;%%},
	Title = {{Relativistic Fluids, Superfluids, Solids and Supersolids from a Coset Construction}},
	Volume = {D89},
	Year = {2014}}

@article{Pufu:2013eda,
	Archiveprefix = {arXiv},
	Author = {Pufu, Silviu S. and Sachdev, Subir},
	Doi = {10.1007/JHEP09(2013)127},
	Eprint = {1303.3006},
	Journal = {JHEP},
	Pages = {127},
	Primaryclass = {hep-th},
	Reportnumber = {MIT-CTP-4447},
	Slaccitation = {%%CITATION = ARXIV:1303.3006;%%},
	Title = {{Monopoles in 2 + 1-dimensional conformal field theories with global U(1) symmetry}},
	Volume = {09},
	Year = {2013}}

@article{PhysRevB.78.214418,
	Author = {Metlitski, Max A. and Hermele, Michael and Senthil, T. and Fisher, Matthew P. A.},
	Doi = {10.1103/PhysRevB.78.214418},
	Issue = {21},
	Journal = {Phys. Rev. B},
	Month = {12},
	Numpages = {10},
	Pages = {214418},
	Publisher = {American Physical Society},
	Title = {Monopoles in ${\text{CP}}^{N\ensuremath{-}1}$ model via the state-operator correspondence},
	Url = {https://link.aps.org/doi/10.1103/PhysRevB.78.214418},
	Volume = {78},
	Year = {2008}}

@article{Dyer:2015zha,
	Archiveprefix = {arXiv},
	Author = {Dyer, Ethan and Mezei, M{\'a}rk and Pufu, Silviu S. and Sachdev, Subir},
	Doi = {10.1007/JHEP03(2016)111, 10.1007/JHEP06(2015)037},
	Eprint = {1504.00368},
	Journal = {JHEP},
	Note = {[Erratum: JHEP03,111(2016)]},
	Pages = {037},
	Primaryclass = {hep-th},
	Reportnumber = {PUPT-2479},
	Slaccitation = {%%CITATION = ARXIV:1504.00368;%%},
	Title = {{Scaling dimensions of monopole operators in the $ \mathbb{C}{\mathrm{\mathbb{P}}}^{N_b-1} $ theory in 2 $+$ 1 dimensions}},
	Volume = {06},
	Year = {2015}}

@article{Banerjee:2017fcx,
	Archiveprefix = {arXiv},
	Author = {Banerjee, Debasish and Chandrasekharan, Shailesh and Orlando, Domenico},
	Doi = {10.1103/PhysRevLett.120.061603},
	Eprint = {1707.00711},
	Journal = {Phys. Rev. Lett.},
	Number = {6},
	Pages = {061603},
	Primaryclass = {hep-lat},
	Slaccitation = {%%CITATION = ARXIV:1707.00711;%%},
	Title = {{Conformal dimensions via large charge expansion}},
	Volume = {120},
	Year = {2018}}

@article{Berenstein:2002jq,
	Archiveprefix = {arXiv},
	Author = {Berenstein, David Eliecer and Maldacena, Juan Martin and Nastase, Horatiu Stefan},
	Doi = {10.1088/1126-6708/2002/04/013},
	Eprint = {hep-th/0202021},
	Journal = {JHEP},
	Pages = {013},
	Primaryclass = {hep-th},
	Slaccitation = {%%CITATION = HEP-TH/0202021;%%},
	Title = {{Strings in flat space and pp waves from N=4 superYang-Mills}},
	Volume = {0204},
	Year = {2002}}

@article{Hellerman:2015nra,
	Archiveprefix = {arXiv},
	Author = {Hellerman, Simeon and Orlando, Domenico and Reffert, Susanne and Watanabe, Masataka},
	Doi = {10.1007/JHEP12(2015)071},
	Eprint = {1505.01537},
	Journal = {JHEP},
	Pages = {071},
	Primaryclass = {hep-th},
	Slaccitation = {%%CITATION = ARXIV:1505.01537;%%},
	Title = {{On the CFT Operator Spectrum at Large Global Charge}},
	Volume = {12},
	Year = {2015}}

@article{Alvarez-Gaume:2016vff,
	Archiveprefix = {arXiv},
	Author = {Alvarez-Gaume, Luis and Loukas, Orestis and Orlando, Domenico and Reffert, Susanne},
	Doi = {10.1007/JHEP04(2017)059},
	Eprint = {1610.04495},
	Journal = {JHEP},
	Pages = {059},
	Primaryclass = {hep-th},
	Reportnumber = {CERN-TH-2016-221},
	Slaccitation = {%%CITATION = ARXIV:1610.04495;%%},
	Title = {{Compensating strong coupling with large charge}},
	Volume = {04},
	Year = {2017}}

@article{Loukas:2017lof,
	Archiveprefix = {arXiv},
	Author = {Loukas, Orestis and Orlando, Domenico and Reffert, Susanne},
	Doi = {10.1007/JHEP10(2017)085},
	Eprint = {1707.00710},
	Journal = {JHEP},
	Pages = {085},
	Primaryclass = {hep-th},
	Slaccitation = {%%CITATION = ARXIV:1707.00710;%%},
	Title = {{Matrix models at large charge}},
	Volume = {10},
	Year = {2017}}

@article{Monin:2016bwf,
	Archiveprefix = {arXiv},
	Author = {Monin, A.},
	Doi = {10.1103/PhysRevD.94.085013},
	Eprint = {1607.06493},
	Journal = {Phys. Rev.},
	Number = {8},
	Pages = {085013},
	Primaryclass = {hep-th},
	Slaccitation = {%%CITATION = ARXIV:1607.06493;%%},
	Title = {{Partition function on spheres: How to use zeta function regularization}},
	Volume = {D94},
	Year = {2016}}

@article{Monin:2016jmo,
	Archiveprefix = {arXiv},
	Author = {Monin, Alexander and Pirtskhalava, David and Rattazzi, Riccardo and Seibold, Fiona K.},
	Doi = {10.1007/JHEP06(2017)011},
	Eprint = {1611.02912},
	Journal = {JHEP},
	Pages = {011},
	Primaryclass = {hep-th},
	Slaccitation = {%%CITATION = ARXIV:1611.02912;%%},
	Title = {{Semiclassics, Goldstone Bosons and CFT data}},
	Volume = {06},
	Year = {2017}}

@article{Hellerman:2017sur,
	Archiveprefix = {arXiv},
	Author = {Hellerman, Simeon and Maeda, Shunsuke},
	Eprint = {1710.07336},
	Primaryclass = {hep-th},
	Reportnumber = {IPMU17-0143, CALT-TH-2017-059},
	Slaccitation = {%%CITATION = ARXIV:1710.07336;%%},
	Title = {{On the Large $R$-charge Expansion in ${\mathcal N} = 2$ Superconformal Field Theories}},
	Year = {2017}}

@article{Komargodski:2012ek,
	Archiveprefix = {arXiv},
	Author = {Komargodski, Zohar and Zhiboedov, Alexander},
	Doi = {10.1007/JHEP11(2013)140},
	Eprint = {1212.4103},
	Journal = {JHEP},
	Pages = {140},
	Primaryclass = {hep-th},
	Slaccitation = {%%CITATION = ARXIV:1212.4103;%%},
	Title = {{Convexity and Liberation at Large Spin}},
	Volume = {11},
	Year = {2013}}

@article{Loukas:2017hic,
	Archiveprefix = {arXiv},
	Author = {Loukas, Orestis},
	Eprint = {1711.07990},
	Primaryclass = {hep-th},
	Slaccitation = {%%CITATION = ARXIV:1711.07990;%%},
	Title = {{A matrix CFT at multiple large charges}},
	Year = {2017}}

@article{Nicolis:2011pv,
	Archiveprefix = {arXiv},
	Author = {Nicolis, Alberto and Piazza, Federico},
	Doi = {10.1007/JHEP06(2012)025},
	Eprint = {1112.5174},
	Journal = {JHEP},
	Pages = {025},
	Primaryclass = {hep-th},
	Slaccitation = {%%CITATION = ARXIV:1112.5174;%%},
	Title = {{Spontaneous Symmetry Probing}},
	Volume = {06},
	Year = {2012}}

@article{Watanabe:2013uya,
	Archiveprefix = {arXiv},
	Author = {Watanabe, Haruki and Brauner, Tom{\'a}{\v s} and Murayama, Hitoshi},
	Doi = {10.1103/PhysRevLett.111.021601},
	Eprint = {1303.1527},
	Journal = {Phys. Rev. Lett.},
	Number = {2},
	Pages = {021601},
	Primaryclass = {hep-th},
	Reportnumber = {BI-TP-2013-02, IPMU13-0056},
	Slaccitation = {%%CITATION = ARXIV:1303.1527;%%},
	Title = {{Massive Nambu-Goldstone Bosons}},
	Volume = {111},
	Year = {2013}}

@article{Nielsen:1975hm,
	Author = {Nielsen, Holger Bech and Chadha, S.},
	Doi = {10.1016/0550-3213(76)90025-0},
	Journal = {Nucl. Phys.},
	Pages = {445-453},
	Reportnumber = {NBI-HE-75-8},
	Slaccitation = {%%CITATION = NUPHA,B105,445;%%},
	Title = {{On How to Count Goldstone Bosons}},
	Volume = {B105},
	Year = {1976}}

@book{Zinn-Justin:572813,
	Address = {Oxford},
	Author = {Zinn-Justin, Jean},
	Publisher = {Clarendon Press},
	Series = {Internat. Ser. Mono. Phys.},
	Title = {{Quantum Field Theory and Critical Phenomena; 4th ed.}},
	Year = {2002}}

@article{Moshe:2003xn,
	Archiveprefix = {arXiv},
	Author = {Moshe, Moshe and Zinn-Justin, Jean},
	Doi = {10.1016/S0370-1573(03)00263-1},
	Eprint = {hep-th/0306133},
	Journal = {Phys. Rept.},
	Pages = {69-228},
	Primaryclass = {hep-th},
	Slaccitation = {%%CITATION = HEP-TH/0306133;%%},
	Title = {{Quantum field theory in the large N limit: A Review}},
	Volume = {385},
	Year = {2003}}

@article{Hellerman:2018xpi,
	Archiveprefix = {arXiv},
	Author = {Hellerman, Simeon and Maeda, Shunsuke and Orlando, Domenico and Reffert, Susanne and Watanabe, Masataka},
	Eprint = {1804.01535},
	Primaryclass = {hep-th},
	Slaccitation = {%%CITATION = ARXIV:1804.01535;%%},
	Title = {{Universal correlation functions in rank 1 SCFTs}},
	Year = {2018}}

@article{delaFuente:2018qwv,
	Archiveprefix = {arXiv},
	Author = {De La Fuente, Anton},
	Doi = {10.1007/JHEP08(2018)041},
	Eprint = {1805.00501},
	Journal = {JHEP},
	Pages = {041},
	Primaryclass = {hep-th},
	Slaccitation = {%%CITATION = ARXIV:1805.00501;%%},
	Title = {{The large charge expansion at large $N$}},
	Volume = {08},
	Year = {2018}}

@book{ZinnJustin:2007zz,
	Author = {Zinn-Justin, Jean},
	Journal = {Oxford, UK: Oxford Univ. Pr. (2007) 452 p},
	Slaccitation = {%%CITATION = INSPIRE-774396;%%},
	Title = {{Phase transitions and renormalization group}},
	Year = {2007}}

@article{McKean:1967xf,
	Author = {McKean, H. P. and Singer, I. M.},
	Journal = {J. Diff. Geom.},
	Pages = {43-69},
	Slaccitation = {%%CITATION = JDGEA,1,43;%%},
	Title = {{Curvature and eigenvalues of the Laplacian}},
	Volume = {1},
	Year = {1967}}

@article{Breitenlohner:1982bm,
	Author = {Breitenlohner, Peter and Freedman, Daniel Z.},
	Doi = {10.1016/0370-2693(82)90643-8},
	Journal = {Phys. Lett.},
	Pages = {197-201},
	Reportnumber = {PRINT-82-0420 (MIT)},
	Slaccitation = {%%CITATION = PHLTA,115B,197;%%},
	Title = {{Positive Energy in anti-De Sitter Backgrounds and Gauged Extended Supergravity}},
	Volume = {115B},
	Year = {1982}}

@article{Appelquist:1982vd,
	Author = {Appelquist, Thomas and Heinz, Ulrich W.},
	Doi = {10.1103/PhysRevD.25.2620},
	Journal = {Phys. Rev.},
	Pages = {2620},
	Reportnumber = {YTP-82-01},
	Slaccitation = {%%CITATION = PHRVA,D25,2620;%%},
	Title = {{Vacuum Stability in Three-dimensional O($N$) Theories}},
	Volume = {D25},
	Year = {1982}}

@article{Appelquist:1981sf,
	Author = {Appelquist, Thomas and Heinz, Ulrich W.},
	Doi = {10.1103/PhysRevD.24.2169},
	Journal = {Phys. Rev.},
	Pages = {2169},
	Reportnumber = {YTP-81-16},
	Slaccitation = {%%CITATION = PHRVA,D24,2169;%%},
	Title = {{Three-Dimensional O(N) Theories At Large Distances}},
	Volume = {D24},
	Year = {1981}}

@article{10.2307/2161165,
	Author = {E. Carletti and G. Monti Bragadin},
	Issn = {00029939, 10886826},
	Journal = {Proceedings of the American Mathematical Society},
	Number = {4},
	Pages = {993--1001},
	Publisher = {American Mathematical Society},
	Title = {On Minakshisundaram-Pleijel Zeta Functions of Spheres},
	Url = {http://www.jstor.org/stable/2161165},
	Volume = {122},
	Year = {1994}}

@article{Vassilevich:2003xt,
	Archiveprefix = {arXiv},
	Author = {Vassilevich, D. V.},
	Doi = {10.1016/j.physrep.2003.09.002},
	Eprint = {hep-th/0306138},
	Journal = {Phys. Rept.},
	Pages = {279-360},
	Primaryclass = {hep-th},
	Slaccitation = {%%CITATION = HEP-TH/0306138;%%},
	Title = {{Heat kernel expansion: User's manual}},
	Volume = {388},
	Year = {2003}}

@inbook{rosenberg_1997,
	Author = {Rosenberg, Steven},
	Booktitle = {The Laplacian on a Riemannian Manifold: An Introduction to Analysis on Manifolds},
	Collection = {London Mathematical Society Student Texts},
	Doi = {10.1017/CBO9780511623783.004},
	Pages = {90--110},
	Place = {Cambridge},
	Publisher = {Cambridge University Press},
	Series = {London Mathematical Society Student Texts},
	Title = {The Construction of the Heat Kernel},
	Year = {1997}}

@inproceedings{stratonovich1957method,
	Author = {Stratonovich, RL},
	Booktitle = {Soviet Physics Doklady},
	Pages = {416},
	Title = {On a method of calculating quantum distribution functions},
	Volume = {2},
	Year = {1957}}

@article{Hubbard:1959,
	Author = {Hubbard, J.},
	Doi = {10.1103/PhysRevLett.3.77},
	Issue = {2},
	Journal = {Phys. Rev. Lett.},
	Month = {Jul},
	Numpages = {0},
	Pages = {77--78},
	Publisher = {American Physical Society},
	Title = {Calculation of Partition Functions},
	Url = {https://link.aps.org/doi/10.1103/PhysRevLett.3.77},
	Volume = {3},
	Year = {1959}}

@article{weyl1911asymptotische,
	Author = {Weyl, Hermann},
	Journal = {Nachrichten von der Gesellschaft der Wissenschaften zu G{\"o}ttingen, Mathematisch-Physikalische Klasse},
	Pages = {110--117},
	Title = {{\"U}ber die asymptotische Verteilung der Eigenwerte},
	Volume = {1911},
	Year = {1911}}

@article{Elizalde:2012zza,
	Author = {Elizalde, E.},
	Doi = {10.1007/978-3-642-29405-1, 10.1007/978-3-540-44757-3},
	Journal = {Lect. Notes Phys.},
	Note = {[Lect. Notes Phys. Monogr.35,1(1995)]},
	Pages = {1-225},
	Slaccitation = {%%CITATION = LNPHA,855,1;%%},
	Title = {{Ten physical applications of spectral zeta functions}},
	Volume = {855},
	Year = {2012}}

@book{kirsten2001spectral,
	Author = {Kirsten, Klaus},
	Publisher = {Chapman and Hall/CRC},
	Title = {Spectral functions in mathematics and physics},
	Year = {2001}}

@article{Banerjee:2019jpw,
	Archiveprefix = {arXiv},
	Author = {Banerjee, Debasish and Chandrasekharan, Shailesh and Orlando, Domenico and Reffert, Susanne},
	Doi = {10.1103/PhysRevLett.123.051603},
	Eprint = {1902.09542},
	Journal = {Phys. Rev. Lett.},
	Number = {5},
	Pages = {051603},
	Primaryclass = {hep-lat},
	Slaccitation = {%%CITATION = ARXIV:1902.09542;%%},
	Title = {{Conformal dimensions in the large charge sectors at the O(4) Wilson-Fisher fixed point}},
	Volume = {123},
	Year = {2019}}

@article{PhysRevD.10.2491,
	Author = {Coleman, Sidney and Jackiw, R. and Politzer, H. D.},
	Doi = {10.1103/PhysRevD.10.2491},
	Issue = {8},
	Journal = {Phys. Rev. D},
	Month = {Oct},
	Numpages = {0},
	Pages = {2491--2499},
	Publisher = {American Physical Society},
	Title = {Spontaneous symmetry breaking in the $\mathrm{O}(N)$ model for large $N$},
	Url = {https://link.aps.org/doi/10.1103/PhysRevD.10.2491},
	Volume = {10},
	Year = {1974}}

@article{guralnik1967broken,
	Author = {Guralnik, Gerald S and Kibble, TWB and Hagen, CR},
	Journal = {Adv. Part. Phys.},
	Number = {PRINT-68-492},
	Pages = {567--708},
	Title = {Broken symmetries and the Goldstone theorem},
	Volume = {2},
	Year = {1967}}

@article{Watanabe:2019xul,
	Archiveprefix = {arXiv},
	Author = {Watanabe, Haruki},
	Eprint = {1904.00569},
	Primaryclass = {cond-mat.other},
	Slaccitation = {%%CITATION = ARXIV:1904.00569;%%},
	Title = {{Formula for the Number of Nambu-Goldstone Modes}},
	Year = {2019}}

@article{Brauner:2010wm,
	Archiveprefix = {arXiv},
	Author = {Brauner, Tomas},
	Doi = {10.3390/sym2020609},
	Eprint = {1001.5212},
	Journal = {Symmetry},
	Pages = {609-657},
	Primaryclass = {hep-th},
	Slaccitation = {%%CITATION = ARXIV:1001.5212;%%},
	Title = {{Spontaneous Symmetry Breaking and Nambu-Goldstone Bosons in Quantum Many-Body Systems}},
	Volume = {2},
	Year = {2010}}

@article{Yaffe:1981vf,
	Author = {Yaffe, Laurence G.},
	Doi = {10.1103/RevModPhys.54.407},
	Journal = {Rev. Mod. Phys.},
	Pages = {407},
	Reportnumber = {CALT-68-843},
	Slaccitation = {%%CITATION = RMPHA,54,407;%%},
	Title = {{Large n Limits as Classical Mechanics}},
	Volume = {54},
	Year = {1982}}

@article{Grassi:2019txd,
	Archiveprefix = {arXiv},
	Author = {Grassi, Alba and Komargodski, Zohar and Tizzano, Luigi},
	Eprint = {1908.10306},
	Primaryclass = {hep-th},
	Slaccitation = {%%CITATION = ARXIV:1908.10306;%%},
	Title = {{Extremal Correlators and Random Matrix Theory}},
	Year = {2019}}

@article{Orlando:2019hte,
	Archiveprefix = {arXiv},
	Author = {Orlando, Domenico and Reffert, Susanne and Sannino, Francesco},
	Doi = {10.1007/JHEP08(2019)164},
	Eprint = {1905.00026},
	Journal = {JHEP},
	Pages = {164},
	Primaryclass = {hep-th},
	Slaccitation = {%%CITATION = ARXIV:1905.00026;%%},
	Title = {{A safe CFT at large charge}},
	Volume = {08},
	Year = {2019}}

@article{Arias-Tamargo:2019xld,
	Archiveprefix = {arXiv},
	Author = {Arias-Tamargo, G. and Rodriguez-Gomez, D. and Russo, J. G.},
	Eprint = {1908.11347},
	Primaryclass = {hep-th},
	Slaccitation = {%%CITATION = ARXIV:1908.11347;%%},
	Title = {{The large charge limit of scalar field theories and the Wilson-Fisher fixed point at $\epsilon=0$}},
	Year = {2019}}

@article{1752533,
	Archiveprefix = {arXiv},
	Author = {Badel, Gil and Cuomo, Gabriel and Monin, Alexander and Rattazzi, Riccardo},
	Eprint = {1909.01269},
	Primaryclass = {hep-th},
        SLACcitation   = "%%CITATION = ARXIV:1909.01269;%%",
	Title = {The Epsilon Expansion Meets Semiclassics},
	Year = {2019}}

@article{1752534,
	Archiveprefix = {arXiv},
	Author = {Watanabe, Masataka},
	Eprint = {1909.01337},
	Primaryclass = {hep-th},
        SLACcitation   = "%%CITATION = ARXIV:1909.01337;%%",
        Title = {Accessing Large Global Charge via the $\epsilon$-Expansion},
	Year = {2019}}

@article{Filothodoros:2016txa,
    author = "Filothodoros, E.G. and Petkou, A.C. and Vlachos, N.D.",
    archivePrefix = "arXiv",
    doi = "10.1103/PhysRevD.95.065029",
    eprint = "1608.07795",
    journal = "Phys.Rev.D",
    number = "6",
    pages = "065029",
    primaryClass = "hep-th",
    title = "$3d$ fermion-boson map with imaginary chemical potential",
    volume = "95",
    year = "2017"
}

@article{Filothodoros:2018pdj,
    author = "Filothodoros, Evangelos G. and Petkou, Anastasios C. and Vlachos, Nicholas D.",
    archivePrefix = "arXiv",
    doi = "10.1016/j.nuclphysb.2019.01.015",
    eprint = "1803.05950",
    journal = "Nucl.Phys.B",
    pages = "195--224",
    primaryClass = "hep-th",
    reportNumber = "CERN-TH-2018-049",
    title = "The fermion-boson map for large $d$",
    volume = "941",
    year = "2019"
}

@article{Romatschke:2019ybu,
    author = "Romatschke, Paul",
    archivePrefix = "arXiv",
    doi = "10.1103/PhysRevLett.122.231603",
    eprint = "1904.09995",
    journal = "Phys.Rev.Lett.",
    number = "23",
    pages = "231603",
    primaryClass = "hep-th",
    title = "Finite-Temperature Conformal Field Theory Results for All Couplings: O(N) Model in 2+1 Dimensions",
    volume = "122",
    year = "2019"
}

@article{DeWolfe:2019etx,
      author         = "DeWolfe, Oliver and Romatschke, Paul",
      title          = "{Strong Coupling Universality at Large N for Pure CFT
                        Thermodynamics in 2+1 dimensions}",
      year           = "2019",
      eprint         = "1905.06355",
      archivePrefix  = "arXiv",
      primaryClass   = "hep-th",
      SLACcitation   = "%%CITATION = ARXIV:1905.06355;%%"
}

@book{Erdelyi:1956asymptotic,
  title={Asymptotic Expansions},
  author={Erd{\'e}lyi, A.},
  isbn={9780486603186},
  series={Dover Books on Mathematics},
  url={https://books.google.ch/books?id=aedk-OHdmNYC},
  year={1956},
  publisher={Dover Publications},
  pages = "46"
}

\end{document}